\documentclass[sigconf,nonacm,pageno]{acmart}

\usepackage[normalem]{ulem}
\usepackage{listings}
\usepackage{xspace}
\usepackage{rotating}
\usepackage{multirow}
\usepackage{xcolor}
\usepackage{makecell}
\usepackage{subfig}
\usepackage[firstpage]{draftwatermark}

\SetWatermarkText{To appear in ASPLOS 2024}
\SetWatermarkAngle{0}
\SetWatermarkFontSize{11pt}
\SetWatermarkLightness{0.2}
\SetWatermarkVerCenter{0.05\pdfpageheight}
\newcommand{\figRef}[1]{\hyperref[fig:#1]{Fig.~\ref{fig:#1}}}
\newcommand{\secRef}[1]{\hyperref[sec:#1]{\S\ref{sec:#1}}}
\newcommand{\tabRef}[1]{\hyperref[tab:#1]{Table~\ref{tab:#1}}}
\newcommand{\codeRef}[1]{\hyperref[lst:#1]{Listing~\ref{lst:#1}}}
\newcommand{\textcode}[1]{\texttt{#1}}

\newcommand{\vcycle}[0]{Vcycle\xspace}
\newcommand{\vcycles}[0]{Vcycles\xspace}

\newcounter{inlineenum}
\renewcommand{\theinlineenum}{\arabic{inlineenum}}
\newenvironment{inlineenum}
  {\unskip\ignorespaces\setcounter{inlineenum}{0}%
   \renewcommand{\item}{\refstepcounter{inlineenum}{(\theinlineenum)~}}}
  {\ignorespacesafterend}

\lstdefinelanguage{pseudo-algo}%
  {classoffset=0,
    keywordstyle=\bfseries,
    morekeywords={
      let, be, arrive-await, barrier, wait, iterations,
      for, while, parallel, nonOpt,
      {==>},
      {-->},
      {<--},
      process
      },
    tabsize=2,
    keywordstyle=[2]\bfseries,
    keywords=[2]{int,uint,List,bool},
    numbers=left,
    stepnumber=1,
    numbersep=5pt,
    classoffset=1,
    classoffset=0,%
    basicstyle=\scriptsize\ttfamily,
    numberstyle=\scriptsize\ttfamily,
    morecomment=[s]{/*}{*/},%
    morecomment=[l]{//},%
    morestring=[b]",
    morestring=[b]',
    captionpos=b,
    frame=tb,
    float=htpb, %prevent splitting on more pages
    xleftmargin=2em,
    xrightmargin=2em
}
\definecolor{commentgreen}{RGB}{2,112,10}
\definecolor{eminence}{RGB}{108,48,130}
\definecolor{weborange}{RGB}{255,165,0}
\definecolor{frenchplum}{RGB}{129,20,83}
\definecolor{ForestGreen}{RGB}{34,139,34}
\definecolor{BrickRed}{RGB}{170,74,68}

\lstdefinelanguage{masm}%
{classoffset=0,
  keywordstyle=\bfseries,
  morekeywords={
    SLL, NOP, PRED, OR, ADD, AND, EXPECT, GST, SLICE, MUX, SEND, MOV, SEQ, \$
    },
  tabsize=2,
  keywordstyle=[2]\bfseries,
  keywords=[2]{int,uint,List,bool},
  numbers=left,
  stepnumber=1,
  numbersep=1.5pt,
  classoffset=1,
  classoffset=0,%
  basicstyle=\scriptsize\ttfamily,
  numberstyle=\scriptsize\ttfamily,
  morecomment=[s]{/*}{*/},%
  morecomment=[l]{//},%
  morestring=[b]",
  morestring=[b]',
  captionpos=b,
  frame=tb,
  xleftmargin=0em,
  xrightmargin=0em,
}
\begin{document}

% \title{Manticore:\\Accelerating RTL Simulation with Fine-Grain Parallelism}
\title{Manticore: Hardware-Accelerated RTL Simulation with Static Bulk-Synchronous Parallelism}

\author{Mahyar Emami}
\email{mahyar.emami@epfl.ch}
\authornote{Equal contributors.}
\affiliation{%
    \institution{EPFL}
    \city{Lausanne}
    \country{Switzerland}
}

\author{Sahand Kashani}
\authornotemark[1]
\email{sahand.kashani@epfl.ch}
\affiliation{%
    \institution{EPFL}
    \city{Lausanne}
    \country{Switzerland}
}

\author{Keisuke Kamahori}
\email{k-kamahori@g.ecc.u-tokyo.ac.jp}
\authornote{Work done during EPFL internship.}
\affiliation{%
    \institution{University of Tokyo}
    \city{Tokyo}
    \country{Japan}
}

\author{Mohammad Sepehr Pourghannad}
\email{mspourghannad@ce.sharif.edu}
\authornotemark[2]
\affiliation{%
    \institution{Sharif University}
    \city{Tehran}
    \country{Iran}
}

\author{Ritik Raj}
\email{ritik_r@ch.iitr.ac.in}
\authornotemark[2]
\affiliation{%
    \institution{Indian Institute of Technology Roorkee}
    \city{Roorkee}
    \country{India}
}

\author{James R. Larus}
\email{james.larus@epfl.ch}
\affiliation{%
    \institution{EPFL}
    \city{Lausanne}
    \country{Switzerland}
}

\renewcommand{\shortauthors}{Emami et al.}

\date{}

\thispagestyle{empty}

\begin{abstract}
  The demise of Moore's Law and Dennard Scaling has revived interest in specialized computer architectures and accelerators.
  Verification and testing of this hardware depend heavily upon cycle-accurate simulation of register-transfer-level (RTL) designs.
  The fastest software RTL simulators can simulate designs at 1--1000~kHz, i.e., more than three orders of magnitude slower than hardware.
  Improved simulators can increase designers' productivity by speeding design iterations and permitting more exhaustive exploration.

  One possibility is to exploit low-level parallelism, as RTL expresses considerable fine-grain concurrency.
  Unfortunately, state-of-the-art RTL simulators often perform best on a single core since modern processors cannot effectively exploit fine-grain parallelism.

  This work presents \emph{Manticore}: a parallel computer designed to accelerate RTL simulation.
  Manticore uses a \emph{static bulk-synchronous parallel} (BSP) execution model to eliminate fine-grain synchronization overhead.
  It relies entirely on a compiler to schedule resources and communication, which is feasible since RTL code contains few divergent execution paths.
  With static scheduling, communication and synchronization no longer incur runtime overhead, making fine-grain parallelism practical.
  Moreover, static scheduling dramatically simplifies processor implementation, significantly increasing the number of cores that fit on a chip.
  Our 225-core FPGA implementation running at 475~MHz outperforms a state-of-the-art RTL simulator running on desktop and server computers in 8 out of 9 benchmarks.
\end{abstract}

\maketitle

\section{Introduction}\label{sec:introduction}

The long-anticipated end of Moore's Law and Dennard Scaling has dramatically increased commercial and academic interest in computational accelerators~\cite{groq_think_fast,catapult_v2,hennessy_new_2019,catapult_v1,tpu_v2,cerebras_stencil,tpu,soda,auto_systolic,HeteroHalide, lstm_accel, fpga_sort,gemmini-dac,cnn_edge}.
As with any hardware artifact, accelerators require many iterations of design, debugging, testing, and software development.
Detailed hardware simulation is at the heart of this activity, and a simulation's turnaround time and throughput can directly affect designer productivity and product quality.

Designers, however, face a dilemma.
Software RTL simulators offer much faster turnaround and better visibility into hardware internals than FPGA prototypes.
Simulation, however, runs far slower than hardware, which can be a bottleneck when simulating a large design, running a long execution, or performing extensive testing.

Since the advent of multicore processors, parallelism has been the preferred approach to improve software performance.
RTL simulation seems to offer many opportunities to follow such an approach.
For example, hardware description languages (HDL) like Verilog or VHDL~\cite{flake_verilog_2020} contain parallel constructs for describing independent hardware components that run in parallel and synchronize only at clock edges.
RTL designs comprise many independent computation tasks.
The designers, however, want circuits to run at high clock frequencies, which limits the number of gates between clock edges.
Consequently, realistic RTL designs comprise many \emph{tiny} tasks.
Modern multicore processors struggle with these fine-grain tasks because synchronization and communication are costly.

This work explores a different approach to increase RTL simulation speed.
\emph{Manticore} is a specialized architecture we designed and built for RTL simulation (i.e., a simulator accelerator).
It uses the bulk-synchronous parallel (BSP~\cite{valiant_bridging_1990}) execution model and static scheduling (i.e., \emph{static BSP}) to eliminate the runtime overheads in communication and synchronization.
Like MIT's Raw machine~\cite{waingold_baring_1997}, Manticore relies entirely on its compiler to schedule resources and communication.
Because RTL code rarely contains long divergent code paths, static scheduling is practical.
The scheduled communication and synchronization run without runtime overhead, so fine-grain interactions among cores are efficient.
In addition, static scheduling  simplifies the Manticore processors, significantly increasing the parallelism possible on a chip.

Manticore's compiler accepts single-clock RTL designs and generates binary code for a Manticore accelerator.
Compilation time is comparable to software compilers, offering software development-like turnaround and a fast simulation rate, especially useful for hours- to day-long simulations.
We prototyped Manticore on an FPGA, and it outperforms Verilator~\cite{snyder_verilator_2020} (the fastest open-source RTL simulator) running on top-of-the-line multicore general-purpose processors despite operating at a fraction of their clock speed.
Hardware-accelerated simulation offers a way out of the dilemma posed above by optimizing ``time to result.''
Small experiments and tests can run on a software simulator with rapid turnaround.
More extensive experiments and tests can run on Manticore, with slightly slower compile times but much faster execution.
And hardware prototypes can be reserved for full-system simulation, operating system bring-up, and software development.

The chief contributions of this work are:
\begin{itemize}
  \item An application of the static BSP execution model to RTL simulation,
  \item The Manticore architecture that employs fine-grain parallelism to simulate RTL,
  \item A compiler that finds parallelism in RTL code and statically schedules it to run effectively on Manticore,
  \item A high-performance FPGA prototype of Manticore,
  \item An extensive evaluation comparing and analyzing the performance of Manticore against state-of-the-art software RTL simulation, and
  \item A demonstration of how to effectively exploit the fine-grain parallelism in RTL simulation.
\end{itemize}

The paper is organized as follows:
\secRef{background} introduces RTL simulation taxonomy and how simulation is performed.
\secRef{static_bsp} presents the static BSP execution model.
\secRef{design} presents Manticore's architecture.
\secRef{micro} discusses a high-performance implementation of Manticore.
\secRef{compiler} presents the compilation techniques used to exploit Manticore's hardware.
\secRef{evaluation} evaluates Manticore's performance and design decisions.
\secRef{discussion} discusses limitations and future directions.
\secRef{related_work} surveys closely related work.
Finally, \secRef{conclusion} concludes.

All of Manticore's components (hardware design, verilog frontend, backend compiler) are publicly available with an MIT license\footnote{\href[]{https://github.com/ManticoreRTL}{https://github.com/ManticoreRTL}}.

\section{Background}\label{sec:background}

RTL simulation can be performed in two ways: \emph{timing-accurate}, or \emph{cycle-accurate}.
Timing-accurate simulation fully models gate delays by timestamping value changes.
By contrast, cycle-accurate simulation captures value changes only at clock edges.
Early in the design process, when logic delays are unknown, cycle-accurate and timing-accurate simulations are similar.

Cycle-accurate simulators are implemented in two ways: \emph{event-driven} or \emph{full-cycle}.
An event-driven simulator observes signals and \emph{dynamically} schedules operations when values change, avoiding unnecessary re-evaluation of unchanging circuit elements.
By contrast, full-cycle simulators use a fixed ahead-of-time schedule to ensure values are computed in the correct order.

This work focuses on full-cycle, cycle-accurate simulation.
Full-cycle is generally faster than event-driven simulation despite the redundant evaluations since the cost of monitoring and scheduling events can outweigh the benefit of avoiding unnecessary execution~\cite{essent_dac}.

\begin{figure}
    \centering
    \includegraphics[width=0.9\linewidth]{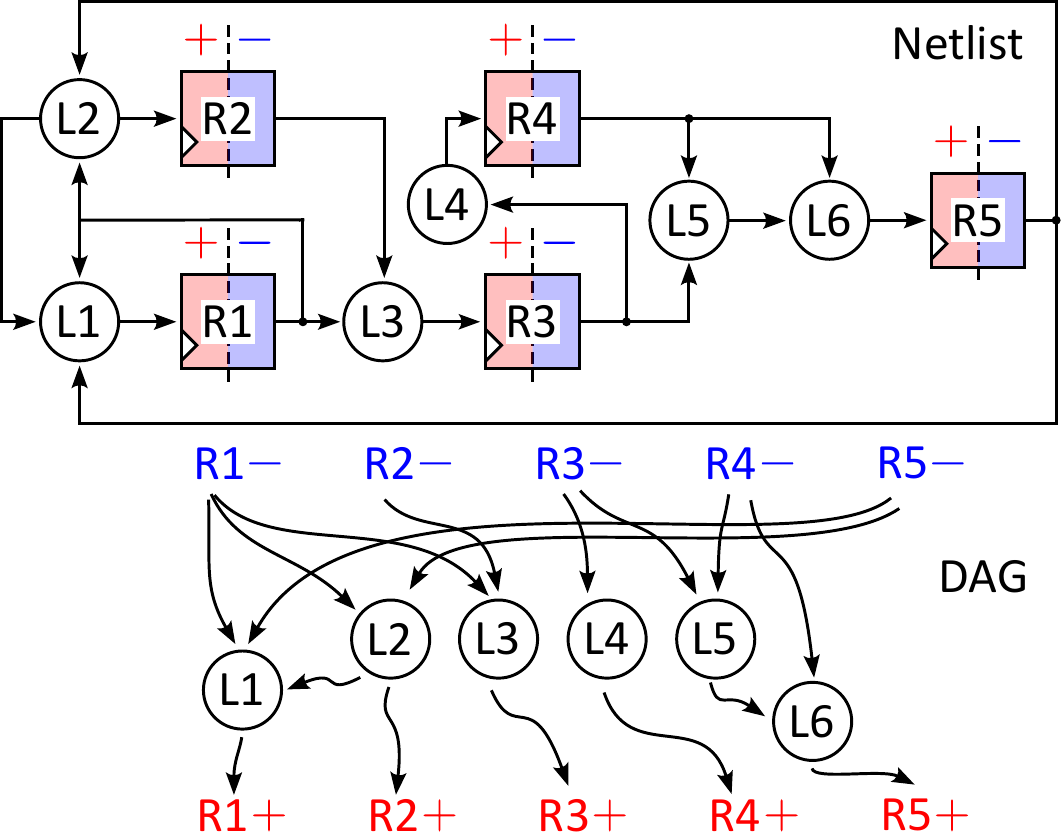}
    \caption{
        An example single-clock netlist (top) and its DAG representation (bottom).
        Circles represent gates and rectangles represent registers.
    }
    \label{fig:parsim}
\end{figure}

\subsection{Full-cycle simulation}

% This section describes Verilator~\cite{snyder_verilator_2020}, our baseline software RTL simulator.
% Verilator is a popular, open-source, full-cycle simulator widely used by academia and industry.
% It performs full-cycle simulation by generating a C++ simulation model, widely believed to run faster than commercial and other open-source simulators~\cite{snyder_verilator_2020}.
% Verilator generates C++ code from an abstract syntax tree (AST) of inlined and RTL code and highly optimizes it with branch prediction hints, short-circuitable branch conditions, and memory prefetch directives.

Hardware description languages model circuits as a \emph{netlist}.
A netlist is a directed graph whose nodes are circuit cells (gates, registers, and memory banks) and whose edges are the wires connecting them.
A netlist graph can be made acyclic by splitting the state nodes (e.g., registers) into a \emph{next} and \emph{current} value.
For example, the top part of~\figRef{parsim} contains a netlist in which circles represent gates and rectangles represent registers.
The corresponding directed acyclic graph (DAG) at the bottom has the \emph{next} and \emph{current} values denoted by \textcode{+} and \textcode{-}, respectively.

Simulating RTL entails evaluating the netlist DAG while respecting precedence relations.
A simulated cycle concludes when all \emph{next} register values have been computed using the \emph{current} register values.
The \emph{current} values are then updated from the \emph{next} values, and the process repeats.
The DAG fully expresses the inherent parallelism of an RTL circuit as an evaluator can traverse independent paths in parallel.

\section{The Static BSP Execution Model}
\label{sec:static_bsp}

\begin{figure}[b]
  \includegraphics[width=\linewidth]{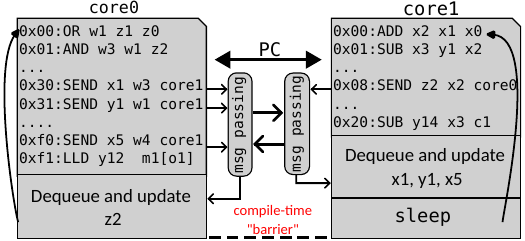}
  \caption{
    The static BSP execution model.
    Each core performs a local computation and then sends its result to the cores that need it for the next computation phase.
    Cores wait (with compiler-inserted \textcode{NOp}s) until all communication completes before starting new computation.
  }
  \label{fig:isa_overview}
\end{figure}

This section describes \emph{Static BSP}, a low-overhead execution model for parallel simulation.
It is inspired by Valiant's bulk-synchronous parallel (BSP) execution model~\cite{valiant_bridging_1990}.
\figRef{isa_overview} depicts the components of static BSP.
Like the original BSP model, ours consists of a system of networked processors that alternate between phases of local computation and cross-processor communication.

\subsection{Runtime Synchronization Freedom}

The original BSP model relied on a \emph{runtime barrier} to synchronize processors at the end of communication (before they start a new computation phase).
Static BSP replaces this barrier with a \emph{compile-time schedule}.
It requires hardware with a \emph{deterministic} interface that permits a compiler to schedule computation and communication.
The compiler uses delay operations (e.g., \textcode{sleep} in~\figRef{isa_overview}) to ensure all processes start the next phase synchronously.

\subsection{Applying Static BSP to RTL Simulation}
\label{sec:static_bsp_rtl}

To parallelize RTL simulation, we partition the netlist DAG (bottom of~\figRef{parsim}) into multiple independent graphs by creating a DAG per sink node.
The computation in each graph consumes multiple \emph{current} register values and produces exactly one value in a \emph{next} register.
The DAGs are independent and can be evaluated in parallel.
Once all DAGs are simulated, the newly computed \emph{next} values become DAG inputs, and a new simulation cycle starts.

RTL simulation can be statically analyzed and scheduled because RTL code rarely has long-lived divergent code paths.
This enables a conservative yet efficient schedule of code paths while maintaining determinism.

In the context of RTL simulation, we call a complete iteration of the computation \emph{and} communication phases a \emph{virtual cycle} (\textbf{\vcycle}).
We do so to distinguish between \emph{RTL cycles} (\vcycle) and the \emph{clock cycles} of the processor that is running the simulation.

% In this model, communications occur at the end of every \vcycle when the newly computed values propagate from producers to consumers.
% This model is similar to Valiant's bulk-synchronous parallel (BSP) model, consisting of alternating cycles of local computation and cross-process communication.
% In practice, separate \emph{current} and \emph{next} registers permit messages to be sent before the barrier, so communication overlaps with computation.
% However, all processors conceptually wait at a barrier for the other processors to compute, send, and receive their values before starting the next \vcycle, so the model is BSP.
%~\figRef{isa_overview} depicts this model.

% \subsection{Runtime Synchronization Freedom}

% Barriers are expensive and not scalable.
% The \emph{runtime} barrier would be unnecessary if the cost of computation and communication were fully known, since a compiler could \emph{statically} introduce delay operations (\textcode{sleep} in~\figRef{isa_overview}) to ensure that no processor starts its next \vcycle before receiving its new values.
% Static scheduling is more scalable than runtime synchronization (i.e., barriers) since the only contention is for the network, not shared synchronization mechanisms.
% However, data-dependent program behavior, dynamic microarchitecture, cache misses, and interference from the operating system and other running tasks make static scheduling impractical in a general-purpose computer.

\section{Manticore Architecture}\label{sec:design}

This section describes Manticore, whose deterministic runtime behavior satisfies the static BSP's requirements.

% Static BSP simplifies hardware implementations as it shifts runtime scheduling decisions typically performed by hardware to a compiler.
% This allows dense, high-frequency implementations that maximize instruction throughput.

\subsection{Key Ideas}

We now list the salient Manticore features that support deterministic behavior.
\begin{itemize}
  \item It consists of simple cores that communicate over a statically-scheduled network-on-chip (\emph{NoC}) (\figRef{manticore_array}).
  \item Cores communicate through \emph{message passing} since shared memory's dynamic performance and communication makes static scheduling difficult.
  \item A compiler schedules how messages are routed between cores to ensure deterministic delivery.
  \item Instructions are \emph{statically partitioned} across cores and stored in fixed-latency, on-chip memories, eliminating frontend stalls so long as each program partition fits a core's instruction memory.
  \item Manticore replaces branches with \emph{predication} and executes all code paths.
  % In RTL simulation, predication can implement the rare and short-lived divergent code efficiently and predictably.
  % Without branches, the compiler statically schedules data hazards.
  \item Each core accesses only the program state stored in its register file and local, fixed-latency scratchpad memory.
  % The large register file practically \emph{eliminates register spills}.
  This memory holds the small (few KiBs) on-chip memories typically found in RTL designs (FIFOs, etc.).
  \item If the program state does not fit in the scratchpads (e.g., processor caches or memories), a \emph{privileged core} accesses an off-chip DRAM memory using a \emph{global stalling} mechanism to ensure all cores and the NoC remain in lockstep.
\end{itemize}
% These architectural features enable a compiler to schedule and control Manticore computation.
% Furthermore, the simplified hardware enables a high clock frequency implementation that scales to hundreds of cores.
% The architecture achieves its two goals of enabling compile-time synchronization and permitting an efficient pipelined hardware implementation that can reach high clock frequencies across hundreds of cores.
% \secRef{micro} contains a detailed description.

% Each instruction can consume up to four operands from the register file.
% Moreover, each core can execute reconfigurable custom instructions to fuse long chains of bitwise logic instructions.

\subsection{Instruction Set}\label{sec:isa}

We briefly describe unconventional aspects of the ISA specific to RTL simulation.

% This section summarizes Manticore's instruction set architecture.
% Manticore has six types of instructions:
% \begin{inlineenum}
% \item standard arithmetic,
% \item custom instructions,
% \item predication,
% \item local memory access,
% \item privileged operations, and
% \item communication.
% \end{inlineenum}

% Its arithmetic instructions include two-operand operations such as addition, subtraction, bitwise logic, and comparison.
% The standard arithmetic instructions take only register operands, but the \textcode{Set rd, imm} instruction updates register \textcode{rd} with a 16-bit immediate value \textcode{imm}.
% A special \textcode{AddCarry} instruction supports efficient simulation of wide additions by reading and producing an \emph{overflow} bit.
% Unlike conventional overflow flags, Manticore exposes an overflow bit in all 2048 registers, each of which can be independently read and written.

Each core supports 32 programmable \emph{functions}, which execute chains of bitwise logic operations with up to four inputs in a single cycle.
E.g., consider the expression:\\
% \begin{equation}
\indent\verb+(a & 0xf) | b | (c & 0x3) | (d ^ 0x1)+\\
% \end{equation}
 with \textcode{a}, \textcode{b}, \textcode{c}, and \textcode{d} being operands\footnote{Taken from picoRV32, a multi-cycle RISC-V processor.}.
A \emph{single} custom instruction replaces these six instructions (see \secRef{cf_insert}).
Custom functions are programmed into a core during boot.

The \textcode{Expect rs1, rs2, eid} instruction raises an exception \textcode{eid} if the values of registers \textcode{rs1} and \textcode{rs2} differ.
Exceptions can invoke services from the host processor (e.g., \textcode{\$display}).
Exceptions, like global memory accesses, stall the execution of all cores and the NoC until they are resolved.
Instructions capable of globally stalling the execution are \emph{privileged} and reserved for a single core, which permits an efficient implementation (see \secRef{global_stall}).

Each core has a scratchpad memory (up to 128KiB) for local load and store operations.
Loads execute unconditionally, but stores are predicated.
Global load and store (predicated) instructions are privileged and access large, off-chip memories using 48-bit addresses.
From the perspective of a compiler, both global and local memory access have the same predictable latencies since the long off-chip latency is masked by stalling all cores and the NoC until a memory access completes.

The producer of a value initiates communication with a \textcode{Send} instruction, which is the only way cores communicate.
The \textcode{Send rt, rs, tid} instruction invoked by a core \textcode{sid} requests target core \textcode{tid} to update its register \textcode{rt} with the value of register \textcode{rs} from core \textcode{sid}.
As~\figRef{isa_overview} illustrates, \textcode{Send}s occur intermixed with computation, but the register updates are delayed until the end of a \vcycle.

\section{Microarchitecture}\label{sec:micro}

This section describes Manticore's microarchitecture and its efficient FPGA implementation.
We prototyped Manticore on a Xilinx UltraScale+ FPGA (an Alveo U200 datacenter FPGA card).
The ideas, however, are general and apply to other FPGAs and ASIC implementations.

\begin{figure}[h]
    \includegraphics[width=\linewidth]{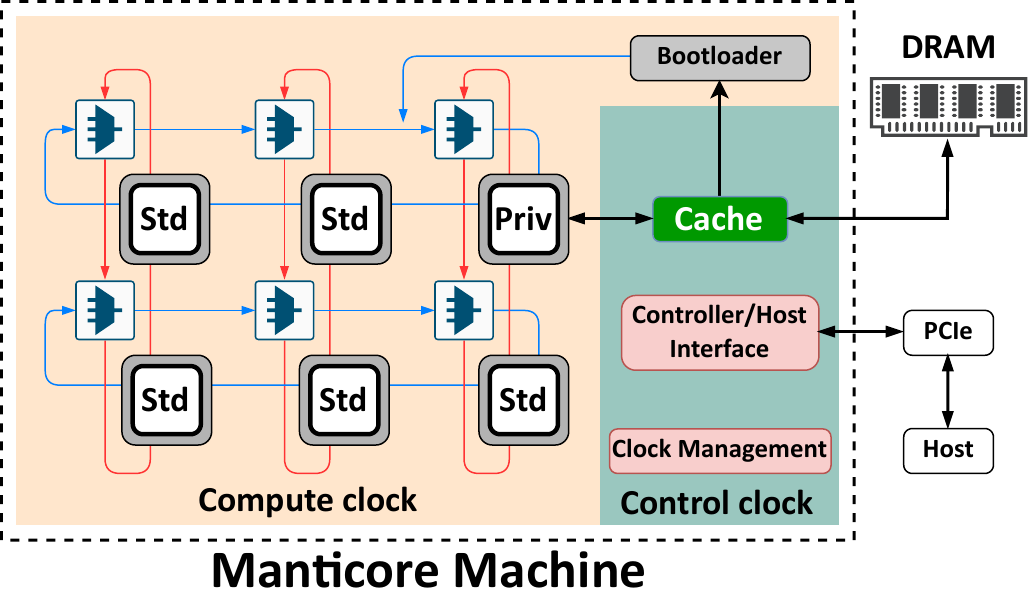}
    \caption{
        A Manticore grid of processors on a uni-directional 2D torus NoC.
        The cores and the NoC reside in the \emph{compute} clock domain, while all other components reside in the \emph{control} clock domain.
        The privileged core is connected to a cache and can access off-chip DRAM.
    }
    \label{fig:manticore_array}
\end{figure}

\figRef{manticore_array} depicts a sample 6-core Manticore grid.
Manticore operates as a accelerator for a \emph{host} (e.g., an x86 processor), which loads programs on Manticore and handles exceptions and termination.
The host has full access to Manticore's DRAM and communicates with Manticore by reading and writing specific registers.
% We will discuss the implementation of each of the building blocks in the following sections.

\subsection{Pipeline Implementation}

Each core is implemented with a simple 14-stage pipeline.
The pipeline is simple because we remove expensive bookkeeping logic (e.g., interlocks and scoreboards) and delegate their function to the compiler.
The logical pipeline consists of the usual five stages: fetch, decode, execute, memory access, and writeback.
Each stage is internally pipelined to achieve a high clock frequency.
A block diagram of the pipeline is in the Appendix (\figRef{manticore_micro}).
RTL code contains structures of various bit widths that are typically narrower than a conventional processor's 32-bit word size.
Manticore uses a 16-bit datapath to match the width of the FPGA's hard DSP units.
This further simplifies the hardware and enables higher clock frequencies.

Instructions are fetched over two cycles from a dedicated instruction memory mapped to a 4096$\times$64 URAM.
URAMs are large 36~KiB on-chip memories.
% All instructions are encoded in 64-bit words.
% We sacrifice space and avoid variable-length encoding to keep decoding simple.

% Instructions with many operands use most of these bits to index the huge register file (e.g., 55 bits for custom instructions).
% Decoding is simple with a set of parallel comparators.
Deep pipelines require a large register file to avoid stalls.
Manticore provides a \emph{2048-entry} register file that exposes all registers to the compiler to avoid expensive hardware renaming logic (similar to the Raw machine~\cite{waingold_baring_1997}).
We implement the register file using BRAMs, configurable 4.5~KiB on-chip memories that support multiple addressing modes.
We use a 2048$\times$17 addressing mode where the lower 16 bits contain the register value, and the most-significant bit contains an \emph{overflow bit} used by wide addition instructions.
The size of the register file requires additional pipelining for reads.
This makes decoding three stages long.
Some instructions can read four values from the register file and write a single result.
This requires four read ports and one write port, which BRAMs do not natively support.
% \footnote{BRAMs and URAMs have one read port and one write port.}.
We use four write-mirrored and identical BRAMs to produce four values simultaneously.

The execute stage consists of two computational units pipelined over four stages.
The ALU handles most standard instructions using a hard FPGA DSP.
The custom function unit (CFU) consists of a small 32$\times$256 memory made of LUTRAMs.
LUTRAMs are FPGA primitives used for shallow memories.
A 1-bit 4-input boolean function is canonically defined by the 16 bits of its truth table.
Manticore's datapath is 16 bits wide, so we extend this idea to a 16-bit truth table using $16 \times 16 = 256$ bits of memory per function.

Scratchpads are mapped to a URAM, with two cycles to access and one cycle to reshape.
We reshape a 4096$\times$64 URAM into a 16384$\times$16 memory by using byte-strobes on the write path and multiplexers on the read path.

\subsection{Network-on-Chip}\label{sec:noc_if}

The cores communicate over a uni-directional torus NoC with buffer-less switching and dimension-ordered routing~\cite{hoplite}.
This design choice reduces routing congestion on the FPGA and supports a high clock frequency.

Switches do not queue messages and immediately route them. % horizontally, vertically, or to the local core.
A switch drops an input message if the target link is busy.
% It consists of stateless pipeline registers and multiplexers.
To avoid data loss, the compiler statically schedules communications.
% To avoid data loss, the compiler ensures timely delivery by scheduling the \textcode{Send} instructions.
Manticore's deterministic execution makes it possible to predict link utilization at each cycle.

Links carry 27 bits of payload, and a few\footnote{Varies based on the grid size, e.g., 8 bits for a 15$\times$15 grid.} bits to specify the target core address.
Messages arriving at a core are queued and received when the core finishes a \vcycle (see~\figRef{isa_overview}).
We use the instruction memory and its unused write port to implement the queue and save resources.
An incoming message encodes an instruction, which is pushed at the end of the instruction memory.
The core subsequently executes it like any other instruction when it reaches it (see~\figRef{manticore_micro} in the Appendix).

% An inbound message is translated to a \textcode{Set rd, imm} instruction on the fly and saved in this memory.
% The \textcode{rd} and \textcode{imm} values both come from the NoC interface.
% The compiler ensures these instructions are written before a core's program counter reaches them.

\subsection{Global Stall}\label{sec:global_stall}

In our current implementation, the privileged core is connected to a 128 KiB direct-mapped, write-allocate, write-back cache backed by a DRAM bank.
The cache is implemented using 4 URAMs.
Accesses to the cache preemptively stall all cores and the NoC until completed, whether that access is a hit or miss.
Therefore, from the compiler's point of view, a global memory access appears to all cores with fixed latency independent of DRAM latency.

Implementing the stall by routing a global signal from the cache to each core would not scale to hundreds of cores.
% However, the routing delays would severely limit the maximum clock frequency, particularly with hundreds of cores.
Instead, we take advantage of the FPGA's clock gating primitives to achieve this functionality.
All parts of Manticore that operate in strict lockstep (the cores and the NoC) reside in the \emph{compute} clock domain.
The rest of the logic that deals with non-determinism reside in the \emph{control} clock domain (see~\figRef{manticore_array}).
The two domains are frequency-matched and phase-aligned.
The logic in the control domain can \emph{halt} or \emph{resume} the compute clock with a global clock buffer as highlighted in~\figRef{manticore_array}.
% by routing the compute clock through a global clock buffer.
% We use one global clock buffer to drive the compute clock.
% The signal that gates the compute clock resides in the control domain.
% \figRef{manticore_array} highlights these two clock domains.

We took great care in implementing the clock gating logic to minimize its effect on scalability.
For instance, there is no logic delay from the clock enable signal to the clock buffer that receives it.
% Furthermore, we manually guide Vivado, Xilinx's FPGA compiler, to select a clock region for routing the global clock signal to minimize skew.
The result is that clock gating logic is nearly independent of the number of cores.

With global clock gating, computation is frozen on a cache request and resumed once completed.
The same mechanism is used to stall the compute domain when an exception occurs so that exceptions are precise.
Control is then transferred to the host machine, and computation resumes at the host's command.

\section{Compiler}\label{sec:compiler}

\begin{figure}[b]
  \centering
  \includegraphics[width=\linewidth]{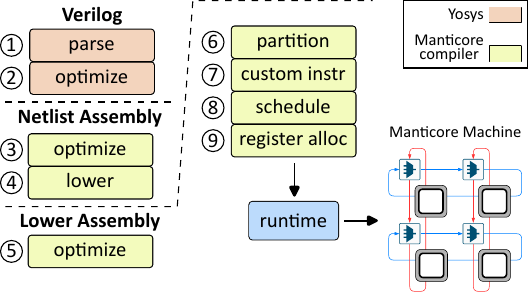}
  \caption{
    Manticore compiler.
    Frontend in red, backend in green.
    A host communicates with the Manticore accelerator through a runtime shown in blue.
  }
  \label{fig:compiler_flow}
\end{figure}

Manticore's hardware was co-designed with its compiler, responsible for extracting parallelism, custom function synthesis, and instruction scheduling.
\figRef{compiler_flow} sketches the compilation process.
The compiler operates on two related intermediate representations (IR): netlist and lower assembly.
Both use static single-assignment and can be interpreted in software.
% The netlist assembly IR has an on-disk textual representation that is the input to the compiler.
% Although human-readable, the netlist assembly is designed to be generated by an HDL frontend.
The lower interpreter is a full-fledged ISA simulator parameterized by the hardware configuration.
We used the interpreters extensively to validate the compiler passes.

We derived our Verilog frontend from Yosys's~\cite{yosys}.
% Since Yosys is inherently a synthesis tool, its Verilog parser lacked the features necessary for simulation.
We extended Yosys to support basic system calls, such as \textcode{\$display} and \textcode{\$stop}, required for simulation.
After parsing the Verilog input, the frontend performs a few optimizations and emits netlist assembly.
Because of the semantics of RTL code, instructions in netlist assembly are unordered and have arbitrary-width operands.

The backend orders the instructions and applies simple optimizations (dead code elimination, constant folding, and common sub-expression elimination).
We then transform the netlist assembly instructions into an equivalent sequence of lower assembly instructions whose operands match Manticore's 16-bit data path.
Initially, the lower assembly is a monolithic sequence of instructions (a single process).
After further optimizations, the compiler partitions the instructions into multiple processes.
The compiler then optimizes each process by fusing chains of bitwise logic instructions into custom instructions.

The final steps of compilation are scheduling and register allocation.
Scheduling ensures that there are no data hazards in the pipeline by inserting \textcode{NOp} instructions to respect data dependencies.
In addition, the \textcode{Send} instructions must be scheduled to ensure timely message delivery.
The compiler then maps virtual registers to machine registers and emits binary code.
The binary is then loaded into Manticore over the NoC by a runtime running on a host x86.
See Appendix~\secRef{runtime_appendix} for details.

The compiler is 18K lines of Scala.
The Yosys Verilog frontend passes are about 2K lines of C++.
The runtime is built on top of the Xilinx runtime library (XRT) with about 800 lines of C++ code.

\subsection{Extracting Parallelism}\label{sec:parallelization}

Partitioning instructions across the cores is the most critical step to achieving good parallel performance.
Despite the absence of runtime synchronization in Manticore, data movement is still costly and excessive communication will limit scalability.
Our parallelization algorithm is aware of this cost and attempts to reduce NoC traffic while distributing work across cores such that each core executes roughly the same number of instructions.

The compiler parallelizes a single monolithic assembly process in two steps:
\begin{inlineenum}
  \item \emph{Split} the monolithic process into a maximal number of tiny processes.
  \item \emph{Merge} the split processes so that the total number of processes does not exceed the number of available cores.
\end{inlineenum}

Splitting follows the approach described in \secRef{static_bsp_rtl} and illustrated in~\figRef{parsim}.
The compiler first creates a DAG representing data dependencies in the monolithic process.
It then uses a backward traversal to partition all nodes reachable from a data sink into an independent smaller process.
In creating the parallel processes, the compiler ensures that instructions that access the same memory region (e.g., an unpacked array in Verilog) end up in the same process to avoid moving large amounts of data every \vcycle.
Additionally, all privileged instructions must execute in the same process.
Partitioning can duplicate DAG nodes across multiple cores, maximizing parallelism at the expense of increased computation.

If we view the maximal set of split processes as a graph whose nodes denote processes and edges denote communication, then merging is a graph partitioning problem.
Existing partitioning tools~\cite{KaHyPar, metis} assume a \emph{linear} cost function, so merging two nodes would add their weight or cost.
However, optimizations such as data sharing and duplicate code elimination make merging non-linear, so we required a heuristic algorithm.

The compiler estimates the execution time of a process as the total number of instructions it executes, including \textcode{Send}s, but excluding the \textcode{NOp}s used to schedule data hazards and received messages.
A vital goal of merging processes is to avoid overloaded cores (i.e., forming stragglers) by equalizing the execution time of all processes.
The compiler iteratively picks two merge candidates that minimize the increase in merged execution time.
It starts from the process with the shortest execution time and merges it with another process with which it communicates.
Intuitively, by starting from the smallest processes and constructing larger ones, we can balance the execution time of the processes and simultaneously reduce communication (hence avoiding network contention).

Merging can continue even after reaching the number of available cores because it can reduce execution time.
For instance, merging processes \textcode{p1} and \textcode{p2} that read a value produced by process \textcode{p3} could lower the execution time of \textcode{p3} because it executes one fewer \textcode{Send} instruction.
Furthermore, since splitting the DAGs may have duplicate code, common sub-expression elimination in a merged process might reduce the number of instructions.

After the merge, the compiler assigns the process that contains privileged instructions to the privileged core.
% \secRef{placement} evaluates other alternatives.

\subsection{Custom Function Synthesis}\label{sec:cf_insert}
Manticore's instructions all have the same latency and programs are branch-free, so shorter programs are faster than longer ones.
Custom function synthesis is the process of collapsing long chains of bitwise logic operations common in RTL simulation into a shallow sequence of 4-input custom functions.
This process is conducted on each partitioned process independently.
Instruction fusion borrows ideas from classical minimum-area, bit-level logic synthesis and applies them to word-level programs.

We start from a process' dependence graph and prune all non-logic vertices.
This leaves us with a set of connected components, each containing only logic operations.
We exhaustively extract all 4-input \emph{maximum fanout-free cones} (MFFC) from each component using cut enumeration~\cite{cong_cut_1999}.
% We start from a process' dependence graph and form a \emph{logic subgraph} by keeping only nodes that represent constants or perform a logic operation.
% The logic subgraph contains multiple disconnected components\footnote{The dependence graph is a single connected component.}.
% We then use cut enumeration~\cite{cong_cut_1999} on each component to exhausitvely extract all its 4-input \emph{maximum fanout-free cones} (MFFC).
% Cut enumeration is an exponential-time algorithm, but is tractable and completes within milliseconds on each component due to their small size.
An MFFC is a tree rooted at a terminal instruction such that no intermediate result is used by an instruction outside the cone. %\footnote{Constants are replciated and folded into MFFCs as they are known at compile-time and do not count as ``inputs''.}.
% MFFCs allow replacing a cone with a single instruction without replicating internal nodes computed by the cone.
Multiple MFFCs can represent the same function and differ only in their representation.
We use logic equivalence checking to group all MFFCs by the function they compute.

Finally, we use a mixed-integer linear programming (MILP) formulation to maximize instruction savings by selecting the best set of non-overlapping MFFCs, while considering that some MFFCs are used at multiple places and yield more savings.
Each MFFC is then replaced with a single custom function.
The MFFCs' truth tables are used to configure each core's CFU at boot time.

\subsection{Scheduling, Routing, and Register Allocation}\label{sec:scheduling}

The compiler uses a simple list-scheduling algorithm to schedule data hazards.
It performs an abstract cycle-accurate simulation of one \vcycle using a model of a core's pipeline and the NoC.
An instruction is scheduled when its predecessors (in the DAG) are scheduled and executed.
Additionally, a \textcode{Send} instruction can be issued only when it will not collide with any other messages on its path.
% We assign a static priority to each core to arbitrate between contending \textcode{Send} instructions from different cores.
If we cannot issue an instruction in a scheduling step, the compiler delays it with a \textcode{NOp} instruction.

Because of the large register file, a simple linear-scan register allocator works well with practically no spills.

Furthermore, we optimize redundant register moves by allocating the same machine register to both the current and next values of an RTL register (e.g., in~\figRef{parsim}, \textcode{+} and \textcode{-} values use the same machine register when possible)~\cite{Wimmer:2010}.

% \subsection{Runtime}
% The runtime is built on top of the Xilinx runtime (XRT) and is responsible for bootloading programs onto a Manticore grid, and handling exceptions or termination.
% Our current implementation is capable of handling simple system calls such as \textcode{\$stop} and \textcode{\$display}.
% We plan to support user-defined Verilog DPI\footnote{Direct programming interface} functions in the future.

\section{Evaluation}\label{sec:evaluation}

This section evaluates Manticore along several dimensions.

\subsection{Fine-Grained Parallel Simulation}\label{sec:bsp_scaling}

We first explore the motivation for a new architecture by studying the limits to fine-grained parallelism in RTL simulation on a general-purpose processor.
We use a simple model of a simulator to find the relationship between simulation speed and computation granularity.
In practice, simulator speed depends on the RTL design and details of the simulator's partitioning, optimization, and runtime.
A fully accurate model is unnecessary if a simplified model offers an upper bound on any system, which we achieve with two simplifications:
\begin{itemize}
  \item Ignore the data transfer among cores and focus exclusively on the synchronization \emph{necessary} to coordinate data movement.
        BSP requires two synchronization points (barriers) per RTL cycle: one at the end of computation and another at the end of communication.
        These are the minimum synchronization needed to simulate an RTL cycle correctly.
        Verilator (our baseline RTL simulator; described in~\secRef{verilator_internals}) also uses two synchronization points as a rendezvous for all tasks at the clock transitions in a cycle.
        % Verilator similarly uses two barriers to coordinate processors at the (two) clock transitions.

  \item As in full-cycle simulation, assume the number of machine instructions required to simulate one RTL cycle is independent of a design's state. This assumption also removes stragglers as a concern.
\end{itemize}

% Let $N$ denote the number of instructions in an RTL cycle, and $P$ denote the number of processing elements (i.e., threads/cores).
% Perfect partitioning assigns $\lfloor N / P \rfloor$ instructions per thread, yielding a $P\times$ speedup.

\subsubsection{First Model}

\begin{lstlisting}[language=pseudo-algo,caption={Model of parallel simulation.},label={lst:barrier_sync_mock}]
// N is the total number of instructions per cycle.
// P is the number of parallel threads.
// B is an arrive-await barrier.
// C is the number of RTL cycles to simulate.
// nonOpt() is an unoptimizable sequence of instructions:
//     a ^= (a+1); b ^= (b+1); c ^= (c+1); d ^= (d+1);
// Independent ops. to avoid read-after-write stalls.
for P parallel threads:
    let localInstr be (N / P / instrInWhileLoop)
    for C iterations:
        // mock computation
        while(localInstr > 0)
            localInstr -= 1
            nonOpt()
        B.wait()
        // mock zero-cost communication
        B.wait()
\end{lstlisting}
\codeRef{barrier_sync_mock} contains the initial model.
The inner loop executes a set of independent arithmetic instructions to model the simulator's computation of an RTL cycle.
The barriers at the end of this computation are necessary to synchronize the communication of newly computed values.
These barriers execute when the model runs and contribute to its runtime cost.
We measure the simulation rate (in kHz) in a strong-scaling experiment that increases the number of threads while keeping the total work constant.
The dashed curves in~\figRef{fake_rate} report the rates on desktop and server x86 systems (details in \tabRef{hw_setup}).

\subsubsection{Second Model}

Model 1 does not fully capture the behavior of a simulator since the \textcode{while} loop has a small instruction footprint that fits in an i-cache.
RTL models are typically larger and incur cache misses.
The fraction of a model that runs on a processor depends on the number of threads; hence, the i-cache performance depends on parallelism.
We fully unroll the \textcode{while} loop to capture this effect.
The differences between the dashed and solid lines in~\figRef{fake_rate} show that simulation speed decreases significantly because of cache pressure.

\subsubsection{Discussion}\label{sec:perf_limit}

This simple model corresponds to Verilator's performance (our baseline RTL simulation; described in \secRef{verilator_internals}).
\secRef{perf_compare} contains measurements of Verilator running nine benchmarks.
\figRef{verilator_scaling} shows that Verilator achieves a maximum speedup of 4$\times$ for the two benchmarks with the largest step size (see~\tabRef{results_all}) and runs slower with multiple threads for smaller benchmarks.

\begin{figure*}
  \centering
  \subfloat{\includegraphics[width=0.25\textwidth]{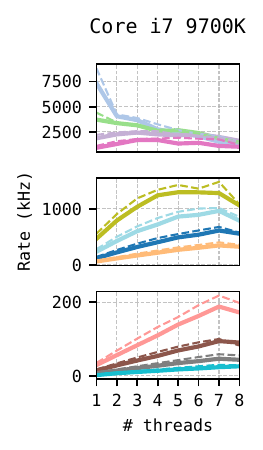}}
  \subfloat{\includegraphics[width=0.75\textwidth]{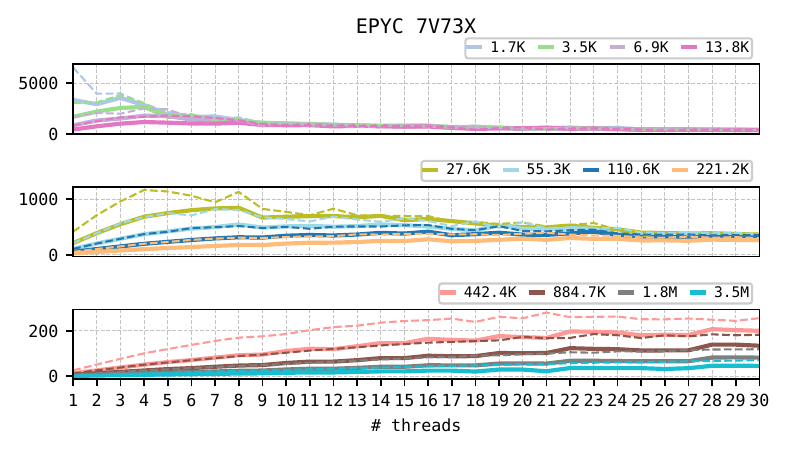}}
  \vspace{0pt}
  \subfloat{
    \resizebox{.85\textwidth}{!}{%
      \begin{tabular}{ l l cccccccccccc  }
        \toprule
        \multicolumn{2}{l}{\bf granularity }      & \textbf{1.7K  } & \textbf{3.5K  } & \textbf{6.9K} & \textbf{13.8K } & \textbf{27.6K } & \textbf{55.3K } & \textbf{110.6K} & \textbf{221.2K} & \textbf{442.4K} & \textbf{884.7K} & \textbf{1.8M} & \textbf{3.5M  }        \\
        \midrule
        \multirow{2}{*}{\rotatebox{90}{\bf i7}}   & {\bf model 1}   & 1.0             & 1.0           & 1.2             & 1.7             & 2.7             & 3.7             & 4.9             & 5.8             & 6.3             & 5.8           & 6.7             & 6.9  \\
                                                  & {\bf model 2}   & 1.0             & 1.0           & 1.3             & 1.8             & 2.8             & 4.2             & 5.3             & 6.0             & 6.5             & 6.6           & 6.6             & 14.5 \\
        \midrule
        \multirow{2}{*}{\rotatebox{90}{\bf epyc}} & {\bf model 1}   & 1.0             & 1.3           & 1.5             & 2.2             & 2.8             & 3.9             & 5.1             & 7.6             & 10.7            & 14.2          & 18.2            & 21.7 \\
                                                  & {\bf model 2}   & 1.1             & 1.6           & 2.1             & 2.8             & 4.0             & 5.2             & 7.9             & 11.5            & 15.9            & 21.4          & 25.6            & 28.4 \\

        \bottomrule
      \end{tabular}
    }
  }\label{tab:fake_speedup}
  \caption{Measured simulated model speed on a desktop (left) and server (right). %
    Dashed lines model only synchronization cost (model 1). Solid lines also include i-cache pressure (model 2). %
    Each curve is labeled by the number of instructions in a simulation step. %
    The table shows the maximum speedup of each model.
  }
  \label{fig:fake_rate}
  % \end{subfigure}
\end{figure*}

Looking in detail, \figRef{fake_rate} identifies three regions of parallel operation:
\begin{itemize}
  \item Small circuits (at most a few thousand instructions) running with very fine-grained parallelism.
        Each clock cycle is a small computation so that serial simulation can reach a few MHz.
        Parallel simulation introduces synchronization every 100--1000 instructions, and its cost causes a steep drop in performance between 1 and 2 processors in the top graphs in~\figRef{fake_rate}.

  \item As the size of a circuit increases, additional processors usually improve performance (middle graphs in~\figRef{fake_rate}).
        In this region, synchronization occurs every 2,000--20,000 instructions.
        Note that the performance benefits are limited, and eventually, the synchronization costs outweigh the benefits of splitting the computation further and the performance decreases.
        This region emphasizes the importance of serial performance; the EPYC processor lags behind the desktop processor, even with its many cores and large caches.

  \item Finally, with hundreds of thousands of instructions in an RTL cycle simulation, parallel execution is beneficial (bottom graphs in~\figRef{fake_rate}) since synchronization is infrequent.
        However, the overall rate is low because each cycle is costly.
        Many cores are needed to push the simulation speed into the 100~kHz range, and the simulation benefits from servers' higher core counts.
\end{itemize}

The figure also displays numerous inflection points where simulation performance decreases with increasing resources.
These inflection points are particularly prominent in fine- and medium-grain simulation.
They occur because additional processors reduce the work-to-synchronization ratio and increase the cost of a barrier.

The table in~\figRef{fake_rate} reports the maximum speedups.
Larger designs offer increased opportunities for speedup.
The second model's speedups are better since its numerator (serial execution) suffers more from i-cache misses than the first model's smaller kernels.
One data point (i7, 3.5M) shows that cache effects can produce super-linear improvement.

Manticore's unconventional design avoids these challenges and can scale its performance over hundreds of cores.
The current prototype allows at most 4096 machine cycles between synchronization points (the instruction memory size).
This puts Manticore in the top region of~\figRef{fake_rate}, where performance scalability is infeasible on a general-purpose computer.
If we are to improve simulation performance through parallelism, adding more cores to general-purpose processors will also result in partitioned workloads that falls in the top region of~\figRef{fake_rate}.

Manticore, however, is limited by its total number of cores and clock frequency.
To match the serial performance of a 4.6--4.9~GHz desktop processor, Manticore must overcome a 10$\times$ reduction in clock speed.
Furthermore, general-purpose computers can execute 1--2.5 instructions-per-cycle (IPC). Manticore's simple processors execute a single instruction per cycle, have a narrower datapath, and support only simple instructions.
Manticore can match the desktop processor's serial performance only if it can achieve a performance improvement of at least 10--25$\times$ by employing parallelism effectively.

\subsection{Manticore FPGA}

We first evaluate the physical design of Manticore's FPGA implementation.
\tabRef{freq} reports the frequency achieved for various Manticore grid sizes.
Smaller grids can operate at \emph{very} high speeds (close to 500MHz).
There is abrupt degradation at the 12$\times$12 grid, explainable by the FPGA's physical layout.
The U200's rectangular floorplan is divided into (1)~a static shell connected to the PCIe bus and (2)~a  user logic region.
The vendor immovably placed the shell at the center-right side of the chip.
User logic has a C-shaped floorplan (Appendix \secRef{floorplanning_appendix} contains die-shots).
With fewer than 160 cores, Manticore fits at the top of the chip, unperturbed by the shell.
Additional cores surround the shell, which complicates timing closure.
We significantly improved the quality of results by guiding the place-and-route tool through the floorplanning of designs with more than 160 cores.
Details are described in the Appendix (\secRef{floorplanning_appendix}).

Each core requires less than 0.021\% of the U200's resources.
The quantity of URAMs limits the number of cores to 398\footnote{Out of 800 available URAMs, the cache requires four.}~\cite{ug1120} (the Appendix contains details).

\begin{table}[h]
  \centering
  \resizebox{0.9\columnwidth}{!}{%
    \begin{tabular}{ c c c c c c c }
      \toprule
      \textbf{Grid}   & \textbf{8$\times$8} & \textbf{10$\times$10} & \textbf{12$\times$12} & \textbf{15$\times$15} & \textbf{16$\times$16} \\
      \midrule
      \textbf{Auto}   & 500                 & 485                   & 480                   & 395                   & 180                   \\
      \textbf{Guided} & --                  & --                    & 500                   & 475                   & 450                   \\
      \bottomrule
    \end{tabular}
  }
  \caption{Clock frequency (MHz) achieved on U200 using automatic and guided floorplanning.}\label{tab:freq}
\end{table}

\subsection{Verilator Simulator}\label{sec:verilator_internals}

We use Verilator~\cite{snyder_verilator_2020} as our baseline RTL simulator.
Verilator is an open-source, full-cycle simulator widely used by academia and industry.
It performs full-cycle simulation and is widely believed to run faster than commercial and other open-source simulators~\cite{snyder_verilator_2020}.
Verilator generates C++ code from an abstract syntax tree of inlined RTL code and highly optimizes it with branch prediction hints, short-circuitable branch conditions, and memory prefetch directives.

Verilator parallelizes RTL simulation by partitioning its DAG into \emph{macro-tasks}, atomic units of work that run asynchronously.
It then combines these tasks into larger units appropriate for multicore processors.
Initially, each DAG node comprises its own macro-task.
Verilator uses Sarkar's algorithm~\cite{Sarkar:87} to increase task granularity by combining macro-tasks that share an edge into a single task.
Combining nodes eliminates the communication of values between cores, but it can increase the critical path of the macro-task graph since the combined nodes execute sequentially.
Verilator's algorithm merges the macro-task that yields the smallest increase in the critical path length.
It does so until it reaches a heuristic threshold for the critical path.
Verilator then statically assigns macro-tasks to a thread pool.
At runtime, a macro-task starts running after its preceding macro-tasks complete execution.
Atomic fetch-and-add operations (i.e., spin-locks) synchronize the macro-tasks, and barriers synchronize the tasks at the clock edges.

Verilator's parallel execution is not BSP since it uses fine-grain synchronization between tasks.
However, in simulating a clock cycle, Verilator uses two synchronization points (final macro-tasks) as a rendezvous for all tasks, similar to the barriers in the model presented above in~\secRef{bsp_scaling} and Manticore.

\subsection{Test Environment}\label{sec:test_env}

We used Verilator v5.006.
Verilator recently added support for multiple clock domains and timing.
Since Manticore does not yet support these, we disable timing in Verilator to avoid penalizing its performance.
We evaluate Verilator's performance on an overclocked desktop and two servers with high core counts.
\tabRef{hw_setup} summarizes the key characteristics of the hardware platforms.

\subsection{Benchmarks}

We evaluate Manticore's performance using nine RTL workloads (the benchmarks are wrapped in simple, assertion-based Verilog test drivers):
\begin{itemize}
  \item \textbf{bc} is a bitcoin miner~\cite{bitcoin}.
  \item \textbf{mm} is a 16$\times$16 integer matrix-matrix multiplier.
  % The two matrices are streamed into the array, one being transposed on the fly.
  \item \textbf{cgra} is a latency-insensitive, coarse-grained reconfigurable array of 64 floating-point processing elements.
        % Data flows through the array using small queues.
        % A separate serial ``bitstream'' is used to configure the grid.
        \item \textbf{vta} is an ML accelerator~\cite{vta}.
        We use a larger\footnote{\textcode{blockIn=64} and \textcode{blockOut=64} instead of 16.} spatial implementation as the default configuration was too small to benefit from hardware acceleration.
        We also divide buffer sizes by 4 to fit in Manticore's scratchpads.
  \item \textbf{rv32r} consists of 16 in-order, pipelined RISC-V processors~\cite{riscv-mini} communicating over a ring network.
        % The processor implementation is based on riscv-mini~\cite{riscv-mini}.
  \item \textbf{jpeg} is a pipelined JPEG decoder~\cite{jpeg}.
  \item \textbf{blur} is a stencil computation accelerator~\cite{stencil:2014}.
  \item \textbf{mc} is a Monte-Carlo simulation stock option price evolution predictor with fixed-point arithmetic~\cite{monte-carlo}.
  % Price evolution is computed using fixed-function pipelines based on Monte-Carlo simulation~\cite{monte-carlo}.
  \item \textbf{noc} is a 2D 4$\times$4 uni-directional torus network-on-chip with wormhole routing and four virtual channels.
\end{itemize}
The benchmarks were sized to ensure their state fit in the Manticore on-chip scratchpads, so the compiler could accurately predict performance.

\begin{table}[h]
    \centering
    \resizebox{1\columnwidth}{!}{%
        \begin{tabular}{ l l l l c }
            \toprule
            {}                & \multicolumn{3}{c}{\bf Baseline (x86)}              & \multirow{2}[2]{*}{\textbf{Manticore}}                                                                            \\
            \cmidrule{2-4}
            {}                & \multicolumn{3}{c}{\bf Verilator v5.006 (Feb 2023)} &                                                                                                                   \\
            \midrule
            \textbf{HW}       & i7-9700K                                            & Xeon 8272CL~\cite{azure_instance_types} & EPYC 7V73X~\cite{azure_instance_types} & \multicolumn{1}{l}{Alveo U200} \\
            \textbf{\# cores} & 8                                                   & 32                                      & 120                                    & \multicolumn{1}{l}{225       } \\
            \textbf{GHz}      & 4.6--4.9                                            & 2.5--3.4                                & 2.2--3.5                               & \multicolumn{1}{l}{0.475     } \\
            \textbf{MiB}      & 14.5                                                & 105.5                                   & 259.6                                  & \multicolumn{1}{l}{18.45     } \\
            \textbf{Date}     & Q4 2018                                             & Q4 2019                                 & Q1 2022                                & \multicolumn{1}{l}{-         } \\
            \bottomrule
        \end{tabular}
    }
    \caption{Hardware platforms. \textbf{\# cores}, \textbf{GHz}, \textbf{MiB}, and \textbf{Date} denote physical core count, clock frequency, SRAM capacity (e.g., cache), and release date.
        SRAM capacities for x86 are computed through \texttt{lscpu}.}
    \label{tab:hw_setup}
\end{table}

\subsection{Performance Comparison}\label{sec:perf_compare}

First, we used the benchmarks to compare the Manticore prototype with Verilator.
We disabled waveform dumps and unnecessary printing and enabled all optimizations in both Verilator (i.e., \textcode{-O3}) and Manticore (e.g., custom functions).
We run each simulation for millions to billions of cycles to capture steady-state performance.
\tabRef{results_all} summarizes the simulation speeds achieved by Manticore and Verilator.

\begin{table}[b]
  \centering
  \setlength{\tabcolsep}{2pt}
  \resizebox{\columnwidth}{!}{%
    \begin{tabular}{l l l c c c c c c c c c  c}
      \toprule
      {}                                                           & {}                                                & {}             & \textbf{vta}                 & \textbf{mc}                   & \textbf{noc}                 & \textbf{mm}                   & \textbf{rv32r}               & \textbf{cgra}                & \textbf{bc}                  & \textbf{blur}                & \textbf{jpeg}              & \multirow{3}{*}{\rotatebox{90}{\textbf{\textit{Geomean}}}} \\
      \cmidrule{4-12}
      \multicolumn{3}{l}{\textbf{\# instr. (k)}}                   & 169                                               & 148            & 88                           & 74                            & 43                           & 38                            & 20                           & 12                           & 3                            &                                                                                                                        \\
      \multicolumn{3}{l}{\textbf{\# cycles}}                       & 1M                                                & 1M             & 1M                           & 1M                            & 1M                           & 1M                            & 2M                           & 5M                           & 1B                           &                                                                                                                        \\
      \midrule
      \multirow{9}[9]{*}{\rotatebox{90}{\textbf{Verilator (kHz)}}} & \multirow{3}[3]{*}{\rotatebox{90}{\textbf{i7}}}   & S              & 41.3                         & 33.9                          & 41.4                         & 43.9                          & 96.6                         & 152.0                        & 599.0                        & 726.7                        & 4246                       &                                                            \\
      {}                                                           & {}                                                & MT             & 160.2                        & 127.2                         & 80.5                         & 83.0                          & 141.8                        & 146.2                        & 354.4                        & 362.0                        & 700.7                      &                                                            \\
      \cmidrule{4-13}
      {}                                                           & {}                                                & {$\times$self} & \textcolor{ForestGreen}{3.9} & \textcolor{ForestGreen}{3.8}  & \textcolor{ForestGreen}{1.9} & \textcolor{ForestGreen}{1.9}  & \textcolor{ForestGreen}{1.5} & \textcolor{BrickRed}{0.97}   & \textcolor{BrickRed}{0.6}    & \textcolor{BrickRed}{0.5}    & \textcolor{BrickRed}{0.2}  & \textbf{\textit{\textcolor{ForestGreen}{1.19}}}            \\
      \cmidrule{3-13}
      {}                                                           & \multirow{3}[3]{*}{\rotatebox{90}{\textbf{xeon}}} & S              & 32.4                         & 26.6                          & 37.1                         & 34.7                          & 97.3                         & 136.8                        & 462.7                        & 532.6                        & 3233                       &                                                            \\
      {}                                                           & {}                                                & MT             & 94.9                         & 68.9                          & 41.5                         & 52.3                          & 73.3                         & 74.3                         & 190.6                        & 186.1                        & 590.6                      &                                                            \\
      \cmidrule{4-13}
      {}                                                           & {}                                                & {$\times$self} & \textcolor{ForestGreen}{2.9} & \textcolor{ForestGreen}{2.6}  & \textcolor{ForestGreen}{1.1} & \textcolor{ForestGreen}{1.5}  & \textcolor{BrickRed}{0.8}    & \textcolor{BrickRed}{0.5}    & \textcolor{BrickRed}{0.4}    & \textcolor{BrickRed}{0.3}    & \textcolor{BrickRed}{0.2}  & \textbf{\textit{\textcolor{BrickRed}{0.79}}}               \\
      \cmidrule{3-13}
      {}                                                           & \multirow{3}[3]{*}{\rotatebox{90}{\textbf{epyc}}} & S              & 32.1                         & 29.7                          & 32.4                         & 31.6                          & 109.2                        & 126.0                        & 550.2                        & 430.5                        & 3627                       &                                                            \\
      {}                                                           & {}                                                & MT             & 146.9                        & 120.8                         & 106.0                        & 95.2                          & 162.7                        & 167.8                        & 370.6                        & 406.9                        & 1239                       &                                                            \\
      \cmidrule{4-13}
      {}                                                           & {}                                                & {$\times$self} & \textcolor{ForestGreen}{4.6} & \textcolor{ForestGreen}{4.1}  & \textcolor{ForestGreen}{3.3} & \textcolor{ForestGreen}{3.0}  & \textcolor{ForestGreen}{1.5} & \textcolor{ForestGreen}{1.3} & \textcolor{BrickRed}{0.7}    & \textcolor{BrickRed}{0.9}    & \textcolor{BrickRed}{0.3}  & \textbf{\textit{\textcolor{ForestGreen}{1.60}}}            \\
      \midrule
      \multirow{7}[7]{*}{\rotatebox{90}{\textbf{Manticore (kHz)}}} & \multicolumn{2}{c}{225-core}                      & 278.1          & 423.0                        & 293.6                         & 567.5                        & 221.0                         & 421.5                        & 1562                         & 1015                         & 214.2                        &                                                                                         \\
      \cmidrule{2-13}
      {}                                                           & \multirow{2}{*}{\rotatebox{90}{\textbf{i7}}}      & {$\times$S}    & \textcolor{ForestGreen}{6.7} & \textcolor{ForestGreen}{12.5} & \textcolor{ForestGreen}{7.1} & \textcolor{ForestGreen}{12.9} & \textcolor{ForestGreen}{2.3} & \textcolor{ForestGreen}{2.8} & \textcolor{ForestGreen}{2.6} & \textcolor{ForestGreen}{1.4} & \textcolor{BrickRed}{0.05} & \textbf{\textit{\textcolor{ForestGreen}{2.75}}}            \\
      {}                                                           & {}                                                & {$\times$MT}   & \textcolor{ForestGreen}{1.7} & \textcolor{ForestGreen}{3.3}  & \textcolor{ForestGreen}{3.6} & \textcolor{ForestGreen}{6.8}  & \textcolor{ForestGreen}{1.6} & \textcolor{ForestGreen}{2.9} & \textcolor{ForestGreen}{4.4} & \textcolor{ForestGreen}{2.8} & \textcolor{BrickRed}{0.31} & \textbf{\textit{\textcolor{ForestGreen}{2.38}}}            \\
      \cmidrule{3-13}
      {}                                                           & \multirow{2}{*}{\rotatebox{90}{\textbf{xeon}}}    & {$\times$S}    & \textcolor{ForestGreen}{8.6} & \textcolor{ForestGreen}{15.9} & \textcolor{ForestGreen}{7.9} & \textcolor{ForestGreen}{16.3} & \textcolor{ForestGreen}{2.3} & \textcolor{ForestGreen}{3.1} & \textcolor{ForestGreen}{3.4} & \textcolor{ForestGreen}{1.9} & \textcolor{BrickRed}{0.07} & \textbf{\textit{\textcolor{ForestGreen}{3.37}}}            \\
      {}                                                           & {}                                                & {$\times$MT}   & \textcolor{ForestGreen}{2.9} & \textcolor{ForestGreen}{6.1}  & \textcolor{ForestGreen}{7.1} & \textcolor{ForestGreen}{10.8} & \textcolor{ForestGreen}{3.0} & \textcolor{ForestGreen}{5.7} & \textcolor{ForestGreen}{8.2} & \textcolor{ForestGreen}{5.5} & \textcolor{BrickRed}{0.36} & \textbf{\textit{\textcolor{ForestGreen}{4.16}}}            \\
      \cmidrule{3-13}
      {}                                                           & \multirow{2}{*}{\rotatebox{90}{\textbf{epyc}}}    & {$\times$S}    & \textcolor{ForestGreen}{8.7} & \textcolor{ForestGreen}{14.2} & \textcolor{ForestGreen}{9.1} & \textcolor{ForestGreen}{18.0} & \textcolor{ForestGreen}{2.0} & \textcolor{ForestGreen}{3.3} & \textcolor{ForestGreen}{2.8} & \textcolor{ForestGreen}{2.4} & \textcolor{BrickRed}{0.06} & \textbf{\textit{\textcolor{ForestGreen}{3.35}}}            \\
      {}                                                           & {}                                                & {$\times$MT}   & \textcolor{ForestGreen}{1.9} & \textcolor{ForestGreen}{3.5}  & \textcolor{ForestGreen}{2.8} & \textcolor{ForestGreen}{6.0}  & \textcolor{ForestGreen}{1.4} & \textcolor{ForestGreen}{2.5} & \textcolor{ForestGreen}{4.2} & \textcolor{ForestGreen}{2.5} & \textcolor{BrickRed}{0.17} & \textbf{\textit{\textcolor{ForestGreen}{2.07}}}            \\

      % \cmidrule{1-13}
      % \multicolumn{2}{c}{\multirow{2}{*}{\rotatebox{0}{\textbf{}}}} & {$\times$S}                                   & \multicolumn{3}{l}{\bf median 3.20} & \multicolumn{3}{l}{\bf geomean 4.41} & \multicolumn{3}{l}{\bf mean 6.20}                                                                                                                                                                                                                                 \\
      % {}                                                                   &                                                & {$\times$MT}                       & \multicolumn{3}{l}{\bf median 3.30}  & \multicolumn{3}{l}{\bf geomean 3.45} & \multicolumn{3}{l}{\bf mean 4.13}                                                                                                                                                                                          \\
      % [2pt]
      \bottomrule
    \end{tabular}
  }
  \caption{%
    Verilator and Manticore simulation performance.
    \textbf{\# instr.} is the average number of x86 instructions (1000s) to simulate one RTL cycle.
    \textbf{\# cycles} is the number of RTL cycles simulated to measure the simulation rate.
    For Verilator, the \textbf{S} and \textbf{MT} rows report the serial and multithreaded simulation performance in kHz.
    \textbf{$\times$self} is multithreaded speedup (wrt serial).
    For Manticore, we report simulation rates on a \textbf{225-core} configuration, along with the speedup relative to the serial (\textbf{$\times$S}) and multithreaded (\textbf{$\times$MT}) runs of Verilator.
  }
  \label{tab:results_all}
\end{table}

\subsubsection{Verilator}

\begin{figure}
  % \subfloat[Core i7-9700K]{\includegraphics[width=\columnwidth]{figures/verilator_threads_i7.pdf}}\\
  \subfloat{\includegraphics[width=\columnwidth]{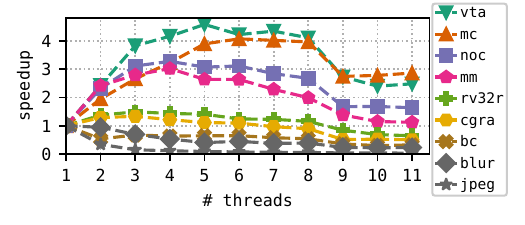}}
  \caption{Verilator parallel scaling on EPYC 7V73X.}
  \label{fig:verilator_scaling}
  % \end{subfigure}
\end{figure}

We report both serial (\textbf{S}) and multithreaded (\textbf{MT}) simulation rates separately for each hardware platform.
Multithreaded Verilator improves performance by up to 3.9$\times$ and 4.6$\times$ on desktop and server processors, respectively.
Multithreading could not improve performance on the smaller benchmarks (e.g., \textbf{bc} and \textbf{jpeg}).
For example, \figRef{verilator_scaling} shows the EPYC processor's scaling trends.
At eight processors, all benchmarks have reached their scalability limit.
Given the number of instructions in each step of the benchmarks, these results accord with the model discussed above in~\secRef{bsp_scaling}.

\subsubsection{Manticore}

The bottom half of~\tabRef{results_all} reports the simulation rates on a 475~MHz 225-core Manticore.
It also reports speedups relative to Verilator's serial (\textbf{$\times$S}) and multithreaded (\textbf{$\times$MT}) performance.
Manticore is consistently faster than Verilator, except for \textbf{jpeg}.
This benchmark has the highest simulation rate in Verilator and the lowest in Manticore.
The \textbf{jpeg} benchmark contains sizeable sequential data dependencies that cannot be parallelized\footnote{Huffman table lookup is the bottleneck.}.
Manticore's slow sequential performance hurts us on this serial benchmark.
Parallelism improves \textbf{jpeg}'s single-core performance by only $\approx$17\%.
This marginal improvement cannot compensate for the single-core disparity between Manticore and x86.

\begin{figure}
  \centering
  \includegraphics[width=\linewidth]{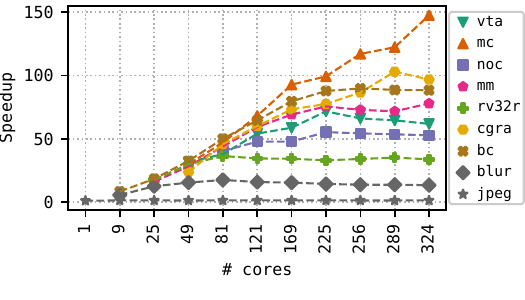}
  \caption{
    Manticore's multicore scaling.
  }
  \label{fig:manticore_scaling}
\end{figure}

\figRef{manticore_scaling} analyzes Manticore's scalability.
The speedup numbers are predicted by Manticore's compiler instead of actual execution, since the compiler can accurately count cycles in the absence of off-chip memory accesses.
The compiler reports a \emph{virtual critical-path length (VCPL)}, the total number of instructions (including \textcode{NOps}) in the slowest core.
VCPL is the number of Manticore machine cycles (i.e., FPGA cycles) required to simulate one RTL cycle.
We consider the single-core VCPL as the baseline, however, on the prototype, single-core execution is, for most benchmarks, impossible since there is not enough space in a single core's instruction memory.

We see that Manticore continues to improve performance as the number of processors increases to 200--300.
Unfortunately, this performance gain through parallelism must be weighed against the single-core/thread performance disparity between an x86 and Manticore.
In other words, a large fraction of the gain goes into making up for the loss in single-core performance.
However, the measurements demonstrate that, with appropriate architectural and compiler support for fine-grain parallelism, we can reach simulation speeds that are \emph{unattainable} on a general-purpose architecture.

Finally, Manticore is not immune to Amdahl's law.
If there is insufficient parallelism in the workload, then Manticore's scaling plateaus.
Depending on the RTL design, this may happen early (\textbf{jpeg}) or late (\textbf{mc}).

\begin{figure}
  \centering
  \includegraphics[width=\linewidth]{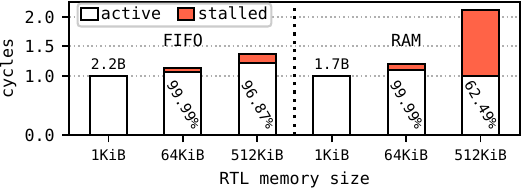}
  \caption{
    Number of machine cycles (lower is better) simulating a FIFO (left) and a RAM (right).
    Numbers are normalized to the cycles needed to run the 1~KiB design.
    Cache hit rate is denoted inside each bar.
  }
  \label{fig:manticore_cache}
\end{figure}

\subsection{Global Stall}\label{sec:eval_stall}

We evaluate the cost of going off-chip with two RTL microbenchmarks running on a 1$\times$1 Manticore grid at 500~MHz: (1)~a FIFO, and (2)~a RAM.
The FIFO and RAM were sized at 1~KiB, 64~KiB, and 512~KiB.
The FIFO reads/writes its memory sequentially, whereas the RAM accesses its memory with pseudo-random addresses (using a simple XOR-shift-128 generator).
Each program runs for 16Mi \vcycles performing a load and store operation per \vcycle.
We use hardware performance counters to log the total number of cycles, stalled cycles, cache hits, and cache misses.
The 1~KiB configuration is a baseline for each microbenchmark since this memory fits in the scratchpad and incurs no global stalls.
The 64~KiB represents a middle point where the state does not fit in the scratchpad but is entirely contained in the 128~KiB cache.
Finally, the 512~KiB configuration corresponds to the scenario where the state is spread between the on-chip cache and off-chip DRAM.
\figRef{manticore_cache} demonstrates that large FIFOs have a high hit rate and are not stall-limited (i.e., FIFOs have excellent spatial locality).
By contrast, randomly accessed RAMs run slower as the number of off-chip accesses increases.
Finally, we observe that cache accesses come at a cost even if they hit since we conservatively stall the execution on every access.

\subsection{Compiler Optimization}\label{sec:compiler_evaluation}

This section evaluates the compiler optimizations.

\subsubsection{Communication-Aware Partitioning}

The balanced partitioning algorithm (B) described in \secRef{parallelization} merges the split processes while keeping communication costs low.
As a baseline, we compare it against communication-oblivious, longest processing-time first partitioning (L) to observe the benefits of modeling communication.
Both algorithms are heuristic and use the same cost estimation method but differ in their merge strategy.
Furthermore, both algorithms are oblivious to the effects of instruction scheduling (after partitioning) as neither accounts for the \textcode{NOp}s for data hazards and NoC contention.

\begin{figure}[!h]
  \centering
  \includegraphics[width=\linewidth]{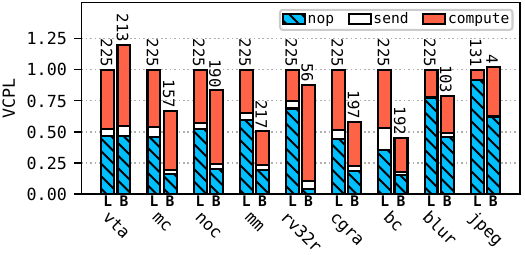}
  \caption{
    Comparison of the communication-oblivious, longest processing-time first algorithm (\emph{L}) and the communication-aware algorithm (\emph{B}) from \secRef{parallelization} for a 15$\times$15 grid.
    VCPL is normalized to the VCPL of L. The numbers above each bar are the number of cores. }
  \label{fig:manticore_parallel}
\end{figure}

\figRef{manticore_parallel} compares the two approaches for a 15$\times$15 Manticore grid, with VPCL normalized to that of L.
We divided the VCPL into the fraction of cycles in the straggler spent computing (\textcode{compute}), sending messages (\textcode{send}), or doing nothing (\textcode{NOp}).
Modeling communication is beneficial as B significantly reduces the overall number of \textcode{Send}s (see~\tabRef{sends}), reduces the number of \textcode{NOp}s in the straggler, and generally outperforms (except for \textbf{vta}) the communication-oblivious algorithm (L) while using fewer cores.
The quality of partitioning significantly affects performance, as evident with \textbf{bc} and \textbf{mm}.

\begin{table}[!h]
  \centering
  \setlength{\tabcolsep}{2pt}
  \resizebox{\columnwidth}{!}{%
    \begin{tabular}{ c c c c c c c c c c }
      \toprule
      {$\times$1000} & \textbf{mm} & \textbf{mc} & \textbf{vta} & \textbf{noc} & \textbf{cgra} & \textbf{rv32r} & \textbf{bc} & \textbf{blur} & \textbf{jpeg} \\
      \midrule
      \textbf{{L}}   & 23.3        & 23.6        & 13.6         & 25.6         & 18.9          & 16.9           & 7.7         & 5.0           & 1.0           \\
      \textbf{{B}}   & 8.5         & 3.9         & 9.8          & 16.6         & 7.4           & 2.8            & 3.1         & 2.7           & 0.1           \\
      \%             & -63.6       & -83.5       & -28.0        & -35.3        & -60.6         & -83.3          & -59.5       & -47.1         & -94.1         \\
      \bottomrule
    \end{tabular}
  }
  \caption{
    \textcode{Send} instructions (1000s) produced by longest processing-time first partitioning (\emph{L}) and balanced partitioning (\emph{B}).
  }
  \label{tab:sends}
\end{table}

\subsubsection{Custom Instructions}

\begin{figure}
  \centering
  \includegraphics[width=\linewidth]{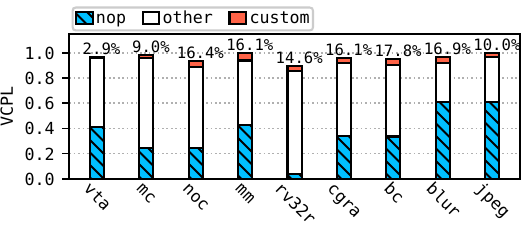}
  \caption{
    Savings in \vcycle due to custom instructions.
    The \vcycle is divided into three instruction types and normalized to not using custom functions.
    The numbers above each bar represent the reduction in non-\textcode{NOp} instructions over all cores.
  }
  \label{fig:manticore_cf}
\end{figure}

We initially proposed custom instructions to compensate for the lack of instruction-level parallelism in Manticore's simple processors by exploiting bit-level parallelism seemingly abundant in RTL.
\figRef{manticore_cf} shows the VCPL of each benchmark normalized to the VCPL without custom instructions.
The VCPL is divided into custom instructions, \textcode{NOps}, and other instructions.
The numbers above each bar show the reduction in the total number of instructions over \emph{all} cores (excluding \textcode{NOps}).
This reduction is 2.9--17.8\%, yet the VCPL (end-to-end) reduction is less than 10\% for all benchmarks.
Custom instructions reduce the \emph{total} instruction count but may not reduce the path length of the straggler (e.g., in \textbf{jpeg}).
Their small benefit comes with a small cost of one BRAM and tens of LUTs per core.
Eliminating the custom instructions would not enable larger Manticore grids since the URAMs are the limiting resource.

\subsubsection{Compile Time}

The Manticore compiler is a prototype built in Scala for robustness.
Its compile times can be several minutes (max. 16m).
By contrast, Verilator compilations usually take less than a minute.

Despite its compilation time, Manticore offers a software development-like experience for longer simulations.
For example, simulating 10B cycles of the \textbf{vta} on Manticore takes about 10 hours and 17 hours on the i7.
In many cases, the extra compilation times are more than compensated by the increased speed.

% Currently, most of the compilation time is spent in parallelization.
% A detailed discussion and breakdown are presented in \secRef{compile_time_appendix}.

\subsection{Cost Analysis}

For completeness, we provide a brief cost analysis using prices from Microsoft Azure.
We estimate the cost of running a few billion simulation cycles in the cloud.
\tabRef{base_cost} shows the Azure instances used in this analysis.
We use the \textbf{D2~v4} instance with two virtual CPUs (vCPU) for serial simulation.
For multithreaded simulation with Verilator, we use the \textbf{D16~v4} instance with sixteen vCPUs.
Furthermore, we also consider the \textbf{HB120rs v3} instance as it lists RTL simulation as a use case.
Renting individual cores on this instance is impossible; therefore, we consider this instance type for only parallel simulation.
Unfortunately, renting an FPGA with a single vCPU in Azure is also impossible.
The smallest instance is the \textbf{NP10s} with one Alveo U250 FPGA board and ten vCPUs, which makes the FPGA instance relatively expensive since we also pay for the unused cores.

\begin{table}[b]
  \setlength{\tabcolsep}{2pt}
  \resizebox{\columnwidth}{!}{%
    \begin{tabular}{ l  c  c}
      \toprule
      {\textbf{Instance}}                                  & \textbf{\$/hour} & \textbf{Simulation} \\
      \midrule
      \textbf{D2 v3}       /  Xeon 8272CL 2$\times$ vCPU   & 0.115            & serial              \\
      \textbf{D16 v4}      /  Xeon 8272CL 16$\times$ vCPU  & 0.92             & multithreaded       \\
      \textbf{HB120rs v3}  /  EPYC 7V73X 120$\times$ vCPU  & 4.68             & multithreaded       \\
      \textbf{NP10s}       /  Alveo U250 + 10$\times$ vCPU & 2.145            & Manticore           \\
      \bottomrule
    \end{tabular}
  }
  \caption{Hourly cost of Microsoft Azure instances~\cite{azure_instance_prices}.}
  \label{tab:base_cost}
\end{table}

All simulations finish in less than an hour for runs shorter than one billion RTL cycles.
With hourly pricing and Verilator's sublinear speedup, serial execution would be the least expensive, followed by multithreaded D16 and Manticore, and finally the HB-series.
However, the cost differences are small (few dollars at most).
More realistically, we consider 1 and 10-billion cycle multiple-hour simulations by estimating the execution time using the simulation rates from~\tabRef{results_all} and then rounding to the next hour (\tabRef{pricing}).
With longer runs, Manticore, in some cases, offers a lower cost than D2 and D16, despite its 2--18$\times$ higher base cost and unused resources.

Far more important, however, is the vast disparity in run duration.
For the ten billion RTL cycle runs, Manticore finishes all of them in a long workday (13 hours).
Multithreaded simulation requires up to two or more full days, while serial simulation can take most of a week.
The productivity gain from several simulation runs per day dwarfs the minor cost savings from using small machines.

\begin{table}[!b]
  \centering
  \setlength{\tabcolsep}{3pt}
  \resizebox{\columnwidth}{!}{%
    \begin{tabular}{l c c c c c c c c c c c}
      \toprule
      {}                                            & {}                                            & {}          & \textbf{vta}  & \textbf{mc}    & \textbf{noc}  & \textbf{mm}   & \textbf{rv32r} & \textbf{cgra} & \textbf{bc}   & \textbf{blur} & \textbf{jpeg} \\
      \midrule
      \multirow{8}{*}{\rotatebox{90}{\textbf{1B}}}  & \multirow{2}{*}{\rotatebox{90}{\textbf{D2}}}  & \textbf{h}  & {\bf 8.58}    & {\bf 10.45}    & { 7.48}       & { 8.00}       & { 2.85}        & { 2.03}       & { 0.60}       & { 0.52}       & { 0.09}       \\
                                                    &                                               & \textbf{\$} & \textbf{1.04} & \textbf{1.27}  & \textbf{0.92} & \textbf{0.92} & \textbf{0.35}  & \textbf{0.35} & \textbf{0.12} & \textbf{0.12} & \textbf{0.12} \\
      \cmidrule{2 - 12}
                                                    & \multirow{2}{*}{\rotatebox{90}{\textbf{D16}}} & \textbf{h}  & { 2.93}       & { 4.03}        & { 6.70}       & { 5.31}       & { 2.85}        & { 2.03}       & { 0.60}       & { 0.52}       & { 0.09}       \\
                                                    &                                               & \textbf{\$} & {2.76}        & {4.60}         & {6.44}        & {5.52}        & {2.76}         & {2.76}        & {0.92}        & {0.92}        & {0.92}        \\
      \cmidrule{2 - 12}
                                                    & \multirow{2}{*}{\rotatebox{90}{\textbf{HB}}}  & \textbf{h}  & { 1.89}       & { 2.30}        & { 2.62}       & { 2.92}       & { 1.71}        & { 1.66}       & { 0.50}       & { 0.52}       & { 0.08}       \\
                                                    &                                               & \textbf{\$} & {9.36}        & {14.04}        & {14.04}       & {14.04}       & {9.36}         & {9.36}        & {4.68}        & {4.68}        & {4.68}        \\
      \cmidrule{2 - 12}
                                                    & \multirow{2}{*}{\rotatebox{90}{\textbf{NP}}}  & \textbf{h}  & { 1.00}       & { 0.66}        & { 0.95}       & { 0.49}       & { 1.26}        & { 0.66}       & { 0.18}       & { 0.27}       & { 1.30}       \\
                                                    &                                               & \textbf{\$} & {2.15}        & {2.15}         & {2.15}        & {2.15}        & {4.29}         & {2.15}        & {2.15}        & {2.15}        & {4.29}        \\
      \midrule
      \multirow{8}{*}{\rotatebox{90}{\textbf{10B}}} & \multirow{2}{*}{\rotatebox{90}{\textbf{D2}}}  & \textbf{h}  & {\bf 85.83}   & {\bf 104.52}   & {\bf 74.77}   & {\bf 79.96}   & {\bf 28.54}    & {\bf 20.30}   & { 6.00}       & { 5.22}       & { 0.86}       \\
                                                    &                                               & \textbf{\$} & \textbf{9.89} & \textbf{12.08} & \textbf{8.62} & \textbf{9.20} & \textbf{3.33}  & \textbf{2.42} & \textbf{0.81} & \textbf{0.69} & \textbf{0.12} \\
      \cmidrule{2 - 12}
                                                    & \multirow{2}{*}{\rotatebox{90}{\textbf{D16}}} & \textbf{h}  & {\bf 29.27}   & {\bf 40.31}    & {\bf 66.97}   & {\bf 53.08}   & {\bf 28.54}    & {\bf 20.30}   & { 6.00}       & { 5.22}       & { 0.86}       \\
                                                    &                                               & \textbf{\$} & {27.60}       & {37.72}        & {61.64}       & {49.68}       & {26.68}        & {19.32}       & {6.44}        & {5.52}        & {0.92}        \\
      \cmidrule{2 - 12}
                                                    & \multirow{2}{*}{\rotatebox{90}{\textbf{HB}}}  & \textbf{h}  & {\bf 18.91}   & {\bf 22.99}    & {\bf 26.21}   & {\bf 29.17}   & {\bf 17.07}    & {\bf 16.56}   & { 5.05}       & { 5.22}       & { 0.77}       \\
                                                    &                                               & \textbf{\$} & {88.92}       & {107.64}       & {126.36}      & {140.40}      & {84.24}        & {79.56}       & {28.08}       & {28.08}       & {4.68}        \\
      \cmidrule{2 - 12}
                                                    & \multirow{2}{*}{\rotatebox{90}{\textbf{NP}}}  & \textbf{h}  & {\bf 9.99}    & { 6.57}        & {\bf 9.46}    & { 4.89}       & {\bf 12.57}    & { 6.59}       & { 1.78}       & { 2.74}       & {\bf 12.97}   \\
                                                    &                                               & \textbf{\$} & {21.45}       & {15.02}        & {21.45}       & {10.72}       & {27.89}        & {15.02}       & {4.29}        & {6.44}        & {27.89}       \\
      \bottomrule
    \end{tabular}
  }
  \caption{%
    Simulation cost using Microsoft Azure prices. %
    Estimated runtime in hours (\textbf{h}) and cost (\textbf{\$}). %
    Embolden hours exceed one workday (8 hours).
    The lowest price is emboldened also.
  }
  \label{tab:pricing}
\end{table}

\section{Limitations and Future Work}\label{sec:discussion}

Manticore explores providing architectural support to accelerate RTL simulation by using fine-grain parallelism.
This paper focused on the technical aspects of our approach, which is still a prototype, not a complete ``tool''.
Nevertheless, \tabRef{results_all} shows a clear performance advantage of the Manticore prototype over a highly optimized software simulator for many examples.
At this maturity level, Manticore is not a replacement for Verilator or other simulators.
Much work is needed to bring Manticore to the same level of usability as Verilator, which has enjoyed more than a decade of active development.

Specifically, Manticore supports only single-clock designs and does not support most of SystemVerilog.
Advanced language features, such as event control, are necessary for a complete simulator, especially for writing complicated test benches.
Accurate timing control (i.e., not cycle-accurate) is incompatible with our approach and would be challenging to retrofit.
However, multiple RTL clock domains could be supported by tracking clock activations independently at each core and conditionally enabling RTL clock domains in all cores.

Waveform debugging is an essential tool in a digital designer's arsenal.
We have an initial design of hardware support for out-of-band waveform collection, but we leave its evaluation for future work.

Current Manticore compile times are longer than a conventional compiler.
This is not an inherent limitation but a byproduct of building a research compiler that allows us to explore alternatives rather than a fast compiler.
Nevertheless, the current compiler offers a faster time-to-result than parallel software simulation for even hour-long simulations.

Currently, most compilation time is spent partitioning the DAG.
Partitioning is necessary for parallel RTL simulation, irrespective of the target hardware (e.g., x86 or Manticore).
The higher degree of parallelism on Manticore makes partitioning slightly more expensive.
We could close the gap between Manticore's and Verilator's compile times with some engineering effort.
In addition, algorithms from research on high-quality, low-complexity partitioning could help in this step~\cite{FM:82,Krishnamurthy:84, Sanchis:89,DSC:94,graph_partitioning_survery:13,graph_partitioning_survery:22}.

Early in the project, we decided to build an FPGA prototype since fabricating an ASIC was not affordable, and simply simulating a Manticore processor would yield less information than constructing an implementation.

The FPGA implementation, however, is limited in its total number of cores and clock frequency.
To match the serial performance of an x86 processor, Manticore must exploit parallelism to overcome a 10--25$\times$ performance loss due to its lower clock speed and IPC.
% To match the serial performance of a 4.6--4.9~GHz desktop processor, Manticore must compensate for the 10$\times$ reduction in clock speed.
% Furthermore, general-purpose computers can achieve 1--2.5 instructions-per-cycle (IPC).
% By contrast, Manticore simple processors are limited to a single instruction per cycle and have narrower datapath and simpler instructions.
% Manticore must utilize parallelism to find a performance improvement of at least 10--25$\times$ to match the desktop processor's single-threaded performance.

In addition to clock frequency, an FPGA has limited SRAM capacity.
Our prototype can simulate up to $\approx$~900k instructions (4096 instructions in each of 225 cores) with about 14.4~MiB SRAM for data and instruction.
Modern accelerator chips contain 100s MiB of SRAM~\cite{groq_maintaining_determinism,groq_think_fast,Graphcore_microbench,darwin}, and hence an ASIC implementation of Manticore could easily avoid these limitations.

\section{Related Work}\label{sec:related_work}

\subsection{FPGA Prototypes and Emulation Platforms}

% Emulators map the RTL to instructions.
% Prototypes map the RTL to gates.

\emph{FPGA prototypes} achieve interactive simulation speeds by mapping RTL circuits to \emph{gates} on an FPGA.
Prototypes can run full software stacks for trillions of clock cycles but require significant engineering effort and lack visibility.
% Commercial prototyping platforms are racks of custom processor grids, costing hundreds of thousands of dollars and simulating at a few MHz.
FireSim~\cite{FireSim} is an open-source FPGA prototyping platform, widely used as an \emph{architectural} simulator for exploring RISC-V designs at datacenter-scale using cloud FPGAs.
\emph{Emulation platforms} are RTL simulators for very large designs~\cite{designing_a_modern_hardware_emulation_platform}.
They provide excellent visibility into hardware state by mapping RTL circuits to \emph{instructions} that run on a processor grid.
Interconnecting multiple custom processor grids (in a rack) generally greatly increases simulation capacity.
However, commercial emulation platforms cost millions of dollars~\cite{emu_compare}.
% It is unclear ho

Although Manticore is implemented on an FPGA, its simulation runs in software (a program running on Manticore) rather than being mapped to an FPGA.
Consequently, Manticore's compile times are a few minutes, whereas FPGA prototypes take hours to days to compile.
Manticore is a first step towards an open-source alternative to commercial emulation platforms.
% It is unclear whether commercial emulation platforms rely on static scheduling.

% While they can be used to simulate arbitrary RTL, it is not their primary focus and come at the cost of significant visibility and engineering effort.
% When full-system simulation at an interactive rate is required, FPGA prototyping offers the most attractive solutions, though at the cost of visibility and huge amounts of engineering effort.
% Multiple prototyping platforms exist.
% While FireSim can simulate general-purpose RTL, it is primarily an \emph{architectural} simulator for exploring RISC-V designs.
% Earlier work like MIT's Virtual Wires~\cite{VirtualWire} project attempt to streamline partitioning a single RTL prototype over multiple FPGAs by virtualizing FPGA pins through multi-pumping.
% Similarly, modern industrial solutions such as Cadence's Protium support mapping upwards of 10 billion gates to 60 FPGAs.

% \subsection{Emulation Platforms}

% Building special hardware for RTL simulation is not new.
% IBM's Yorktown Simulation Engine (YSE)~\cite{YSE} did something similar 40 years ago!
% YSE implements special purpose \emph{logic processors} capable of computing arbitrary 2-bit wide logic functions using table lookups like Manticore's custom functions.
% EDA companies offer similar platforms.
% Cadence's Palladium Z1 and Siemens' Veloce Strato+ are racks of custom processor grids with custom operating systems and compilers offering simulation rates in the range of a few MHz with excellent visibility.
% However, these platforms cost hundreds of thousands of dollars.

\subsection{Parallel RTL simulation}

There is considerable research on accelerating RTL simulation using parallelism, especially GPUs.
Much of this work demonstrates significant speedups relative to commercial \emph{event-driven} simulators.
By contrast, Manticore is a \emph{full-cycle} simulator, so it is not comparable to these systems.
Most of this work focused on reducing the runtime overhead of monitoring value changes, for example, GCS~\cite{Chatterjee:2009, GCS, GCS:journal} or Qian and Deng~\cite{Qian:2011}.
They improved simulation rates by orders of magnitude, to $\approx$~5--37~kHz, over single-threaded commercial simulators.
Manticore operates at rates exceeding 115~kHz (see~\tabRef{results_all}).

% We provide a short survey of the most relevant (and recent) work for completeness regardless.

% GCS~\cite{Chatterjee:2009, GCS, GCS:journal} is a hybrid event-driven GPU-accelerated event driven simulator that levelizes logic gates in coarser macro-gates for decreased runtime overhead associated with monitoring nets.
% It reports orders of magnitude faster runtime (5~kHz simulation rate) compared to single-threaded commercial simulators\footnote{Manticore's typical performance is above 150~kHz, see~\figRef{manticore_scaling}}.

%% Numbers reported by GCS:
%% cycles, GPU time in second,
% >> import numpy as np
% >> gcs = [(12889495, 2567),
%  (13423608, 7781),
%  (1000000, 2578),
%  (10000000, 25973),
%  (2983674, 599),
%  (1967155, 397),
%  (10000001, 3935),
%  (1074702, 6077),
%  (1074702, 8229),
%  (1074702, 10983)]
% >> rate = [x / y * 1e-3 for (x, y) in gcs]
% >> np.max(rate)
% 5.021229061160888

% Qian and Deng~\cite{Qian:2011} propose an event-driven GPU-accelerated RTL simulation based on the Chandy-Misra-Bryant~\cite{Bryant:1977, Chandy:1976} distributed simulation algorithm.
% They show up to 50$\times$ faster simulation versus a single-threaded commercial simulator though on a low-end desktop CPU reaching a simulation rate of 37kHz.
% they simulate for 1e5 cycles and report the following numbers
% qian = [4.019, 10.21, 3.54, 2.67]
% rate = [24.88181139586962, 9.79431929480901, 28.248587570621467, 37.453183520599254]

% SystemC paper also simulates for 1e5 cycles
% sc = [3.916, 1.37, 0.277, 0.276, 0.146]
% scrate = []
RTLFlow~\cite{RTLFlow} is a GPU-accelerated RTL simulator that exploits stimulus-level parallelism to speed up simulation by running many independent simulations on a GPU.
RTLFlow improves execution speed by up to 40$\times$ over Verilator for many stimuli, but it runs an order magnitude slower than Verilator with a single stimulus.
Manticore is faster than Verilator with a single stimulus.

% SCGPSim~\cite{Nanjundappa:2010} accelerates SystemC~\cite{systemc} simulation on GPUs.
% % SystemC is a C++ library for event-driven simulation of circuit mo#dels which conventionally relies on userspace cooperative threads (mapped to a single kernel thread) for modeling concurrent processes.
% They propose a new thread model for GPU execution and report up to 100x faster simulation time compared to a \emph{laptop CPUs} with microbenchmarks.

Zhang~\cite{zhang2020opportunities} called for a renewal in GPU-accelerated RTL simulation research by leveraging recent advances in GPU-compute APIs designed for machine learning.

Nexus~\cite{nexus_emulation:22} is an FPGA-based open-source emulation platform. It uses an array of dynamically scheduled logic processors with an 8-bit data.
Similar to Manticore, Nexus leverages FPGA LUTs to accelerate RTL simulation.
However, unlike Manticore, Nexus does not have a standard ALU and hence \emph{all} logic operations are emulated using the LUTs.
At the time of this writing, Nexus does not have a functional compiler.
Therefore we are not able to provide a quantitative comparison between Manticore and Nexus.

DyVe~\cite{Strauch19} is an event-based, cycle-accurate RTL simulator running on a custom array of many-core SoCs linked with a central FPGA.
DyVe partitions the circuit graph by its primary outputs, then incrementally merges program regions that share the largest number of inputs.
% This process repeats until the number of partitions is equal to the number of processors (unlike Manticore which may use fewer processors than those available if it results in a shorter schedule).
DyVe's performance numbers are based on whether a target's simulation code fits its processors' L1, L2, or SDRAM memories, so direct comparisons are impossible.

\subsection{Sequential RTL Simulation}

Most efforts in improving RTL simulation on CPUs focused on reducing the runtime overhead of event-driven simulation.
ESSENT~\cite{essent_dac,essent_iccad} is a cycle-accurate simulator that employs a coarsened, conditional, singular, static (CCSS) execution model~\cite{essent_dac}.
CCSS is a novel, hybrid approach that minimizes the overhead of runtime checks in event-driven simulation, especially in the presence of low activity factors.
ESSENT is single-threaded and accelerates simulation of RISC-V cores (CPUs have low activity factors) by 1.5--11.5$\times$ over Verilator.
However, it is not clear how ESSENT performs with spatial designs that exhibit high activity factors, especially since it is single threaded~\cite{essent_micro}.
Manticore's performance is independent of a design's activity factor.

Cuttlesim~\cite{Cuttlesim} is a cycle-accurate simulator for Kôika~\cite{Koika}, a rule-based HDL derived from Bluespec Verilog~\cite{Bluespec}.
Cuttlesim uses the high-level semantics of Kôika to generate C++ code optimized for sequential performance.
It reports 2--3$\times$ faster simulation than the equivalent RTL code running serial Verilator.

% \subsection{Fine-grained parallelism}

\subsection{Deterministic Acceleration}

Manticore's design philosophy is similar to VLIW processors and other Raw machines~\cite{waingold_baring_1997}.
A more recent example is Groq's ML accelerator~\cite{groq_maintaining_determinism, groq_think_fast}.
The Groq chip has deterministic hardware datapaths that enable precise reasoning and control by software.
Like RTL simulation, machine learning exhibits rare long-lived divergent code paths, which makes static scheduling feasible.

\section{Conclusion}\label{sec:conclusion}

RTL simulation is an essential aspect of hardware design, and improved simulation offers many benefits to hardware designers.
Currently, a designer must choose between an FPGA prototype's long compile times and fast execution or an RTL simulator's fast turnaround and slow speed.
RTL simulation is slow because even state-of-the-art simulators fail to improve their performance by exploiting the abundant fine-grained parallelism in RTL circuits due to the high communication and synchronization cost of modern processors.

This work presented Manticore, a prototype, hardware-accelerated RTL simulator.
Manticore's processors expose a \emph{deterministic} hardware interface that allows a compiler to \emph{statically} schedule programs across hundreds of simple cores.
This approach eliminates the costly runtime overhead of synchronization, which enables efficient parallel simulation of RTL circuits and allows hundreds of cores to fit on a single chip.
Our prototype FPGA Manticore implementation consistently achieves better performance over a state-of-the-art software RTL simulator.
The Manticore system demonstrates the actual performance benefits of exploiting fine-grained parallelism in RTL code to accelerate simulation.
Its higher speed allows several long simulations per day, as opposed to several per week on a conventional computer, with a concomitant improvement in developer productivity.

\bibliographystyle{plain}
\bibliography{references}

\clearpage
\appendix

\section{Appendix}\label{sec:appendix}

This appendix contains supplementary material that elaborates on points raised in the body of this paper.

\subsection{Verilator's scaling}

\begin{figure}
  \centering
  \includegraphics[width=\linewidth]{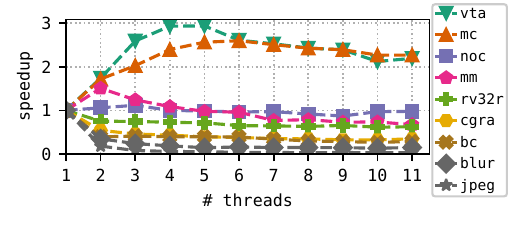}
  \caption{Verilator's parallel scaling on Xeon 8272CL.}
  \label{fig:verilator_threads_xeon}
\end{figure}

\begin{figure}
  \centering
  \includegraphics[width=\linewidth]{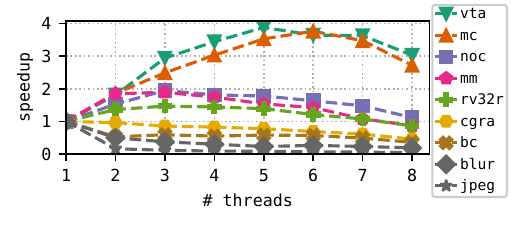}
  \caption{Verilator's parallel scaling on Core i7-9700K.}
  \label{fig:verilator_threads_i7}
\end{figure}

\figRef{verilator_threads_xeon} and~\figRef{verilator_threads_i7} demonstrate Verilator's self-relative speedup on the Intel Xeon 8272CL and Core i7-9700K respectively.

\subsection{Microarchitecture}

\begin{figure*}
  \centering
  \includegraphics[height=0.93\textheight]{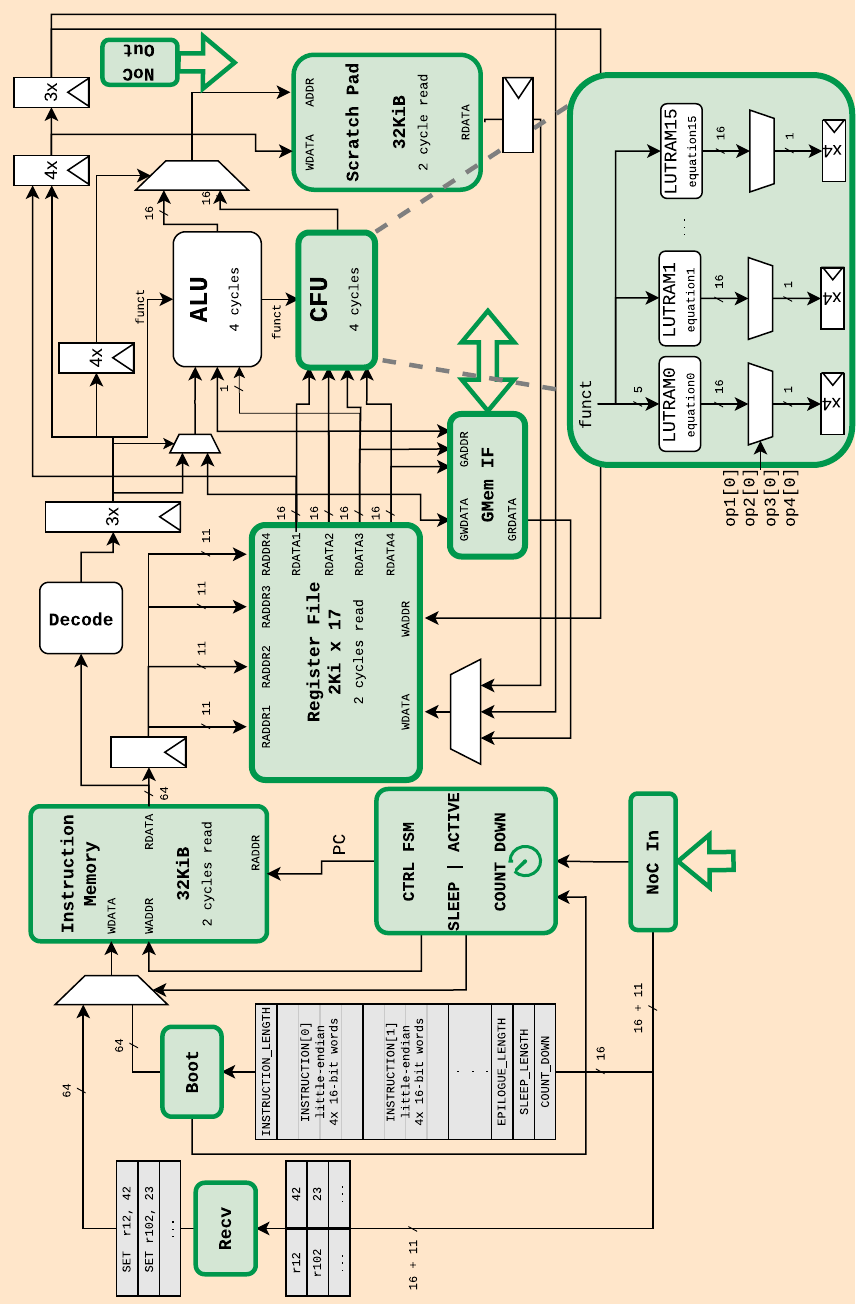}
  \caption{
    Microarchitecture of a core in Manticore's processor grid. Details are omitted for legibility.
  }
  \label{fig:manticore_micro}
\end{figure*}

\figRef{manticore_micro} outlines the pipelined implementation of one core.
The pipeline is 14 stages deep and is logically divided into the typical five functions: fetch, decode, execute, memory access, and writeback.
The CFU is implemented as 16 LUTRAMs.

The pipeline's frontend (right side of~\figRef{manticore_micro}) is responsible for receiving messages during simulation.
Each message is translated on the fly to a \textcode{SET} instruction and is written into instruction memory.
A \textcode{SET} instruction updates a register with an immediate value.
A state machine controls the execution of the pipeline, which is kept in strict lock-step with all other cores.
The compiler inserts sleep instructions to coordinate communication between cores.
An additional state machine handles incoming messages from the bootloader (see \secRef{bootloader_appendix}) and fills the instruction memory before the simulation starts.

\subsection{Runtime}\label{sec:runtime_appendix}

Manticore's runtime is a program running on the host processor that takes a binary generated by the compiler and runs it on the Manticore hardware accelerator connected to the host.
The runtime copies the program binary into FPGA DRAM, then instructs the hardware bootloader (see~\figRef{manticore_array}) to copy the program into the local instruction memories.
While the code executes, the runtime continuously polls the hardware state registers to handle exceptions or terminate execution.

\subsubsection{Bootloader}\label{sec:bootloader_appendix}
Bootloading starts with a soft reset that brings all cores to a \emph{boot state}.
The soft reset only changes a few state registers in each core; it does not reset the register files or the scratchpads.

Cores continuously push \textcode{NOps} through their pipelines and snoop the NoC for instructions when in the boot state.
A hardware bootloader (see~\figRef{manticore_array}) module reads the program binary from DRAM and streams the instructions to each core in sequence.
Cores store the incoming instructions in their instruction memory.
The cores then wait for a message from the NoC that contains a core-specific countdown value.
At that point, the cores initialize a local timer with their specific countdown and start execution when the timer counts down to 0.
The countdown starts all cores simultaneously despite the non-deterministic time required to read a program binary from DRAM.

\figRef{manticore_micro} shows the instruction stream format that a core receives.
This stream consists of a header that encodes the number of instructions in a program and a footer comprised of three words:
\begin{enumerate}
  \item \verb+EPILOGUE_LENGTH+ denotes the total number of messages the core expects to receive at every \vcycle.
  \item \verb+SLEEP_LENGTH+ denotes the sleep period length (see \textcode{sleep} in~\figRef{isa_overview}).
  \item \verb+COUNT_DOWN+ is the last word received that initiates a countdown to the start.
\end{enumerate}

% \subsubsection{Initialization}

% Before booting the simulation program, the runtime first bootloads and runs a few \emph{initializers}.
% An initializer is a Manticore program that initializes the registers and scratchpads or configures the CFUs.
% There is typically more than one initializer since the scratchpads are larger than the instruction memory and initializing them requires running multiple programs.
% The compiler generates these initializers, and the runtime bootloads them into Manticore, runs them one by one, then bootloads the actual program to perform the simulation.

\subsubsection{Exceptions}

Manticore's hardware design and execution model make it possible to pause the execution, perform some computation on the host, and resume the execution on the FPGA.
An example can illustrate this.
Consider the Verilog statement:

\verb+if (count != 0) $display("got %d", count);+.

\noindent This statement is executed as a global store instruction predicated by the condition \verb+count != 0+ that stores \verb+count+ to the global memory.
The \verb+$display+ system call is translated to an \textcode{EXPECT} instruction that throws an exception when \textcode{count} is non-zero.
When Manticore raises the exception, the grid stalls globally, and the host flushes the cache and reads the value of \textcode{count} from the FPGA DDR memory.
The runtime prints this value for the user.
Currently, our compiler only supports basic Verilog system calls.
We plan to support arbitrary DPI (Verilog Direct Programming Interface) calls through this mechanism.
However, crossing the host-device boundary is very expensive and should be avoided as much as possible.

\subsection{Verilog to Assembly}

\codeRef{verilog_example} shows a simple Verilog module that prints a message to the console at every cycle.
The equivalent representation in Manticore assembly using two processes is given in~\codeRef{masm_example}.
This code for each process is repeatedly executed until some exception is raised by the \textcode{EXP} instructions.
When that happens, execution on Manticore freezes until the host processor take the necessary action, for instance print the message or terminate simulation (e.g., \textcode{\$finish})
Each process performs and implicit loop that is padded with extra \textcode{nops} in way that all processes jump back to their starting position at the same time.

\subsection{Floorplanning}\label{sec:floorplanning_appendix}

\begin{figure*}
  \centering
  \includegraphics{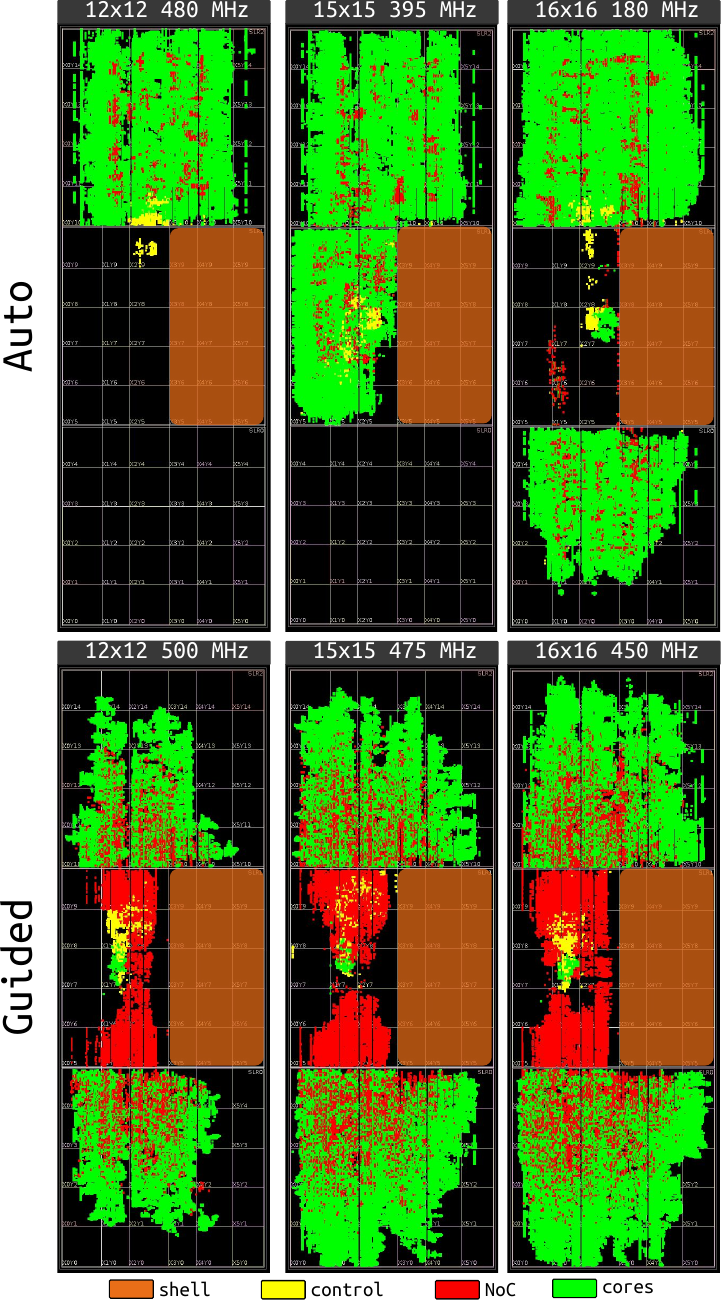}
  \caption{
    Floorplan of Manticore's physical implementation on U200.
    Vivado's automatic floorplanning is at the top and our guided floorplanning is at the bottom.
    The cores are colored in green, the NoC in red, and the control clock domain in yellow.
    The fixed shell is marked as an orange box in each floorplan.
    The clock speed is considerably higher with the guided floorplans.
  }
  \label{fig:floorplan}
\end{figure*}

The U200 is a large \emph{multi-die} FPGA that contains three SLRs\footnote{Super Logic Regions, Xilinx terminology for a single die.}.
Inter-SLR connections are significantly more costly than intra-SLR ones.
While Vivado can find an efficient floorplan of Manticore's torus structure when the full design fits in a single SLR, it fails to do so when SLR crossings are necessary (see~\figRef{floorplan}).

We get around this limitation by using a floorplanning script to guide Vivado.
Cores do not directly access the shell and communicate only with their corresponding switch through a pipelined path.
Therefore, cores do not need to respect a torus topology, so we spread half the cores in the top SLR and the other half in the bottom SLR.
However, the NoC switches must be connected in a torus structure, so we constrain them to the narrow rectangular region in the central SLR.
We constrain Vivado to use a set of dedicated hard registers available for crossing SLRs for each core-to-switch connection.
We also co-locate the privileged core, cache, bootloader, and clock control logic in the central SLR as they access the shell.
Finally, we minimize clock skew between the compute and control clock domains by assigning the clock buffers and clock roots to the same clock region.
These optimizations enable a 15$\times$15 grid to run at 475 MHz.

\subsection{Compile Time Analysis}\label{sec:compile_time_appendix}

\begin{figure}
  \centering
  \includegraphics[width=\linewidth]{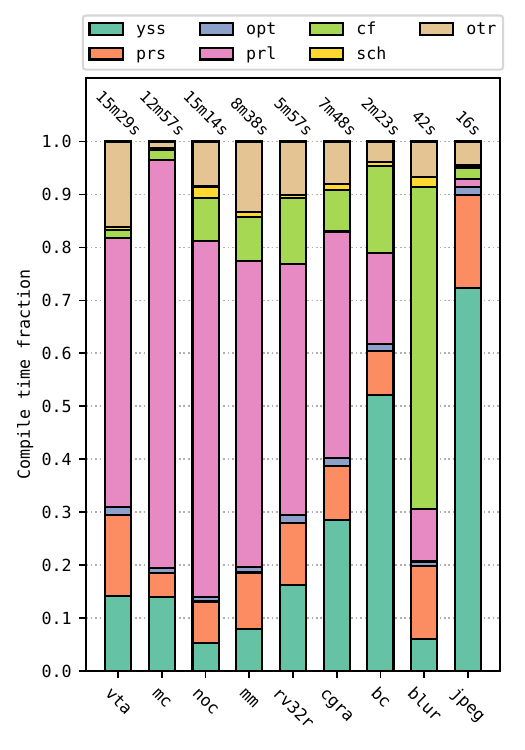}
  \caption{
    Breakdown of compilation time:
    Yosys (\textcode{yss}), assembly parsing (\textcode{prs}), basic optimizations (\textcode{opt}), parallelization (\textcode{prl}), custom function extraction (\textcode{cf}), scheduling (\textcode{sch}), others (\textcode{otr}).
  }
  \label{fig:compile_time}
\end{figure}

\tabRef{compile_time} reports the compilation times of each benchmark for both Manticore and Verilator (single-thread and multithread).
Manticore's compiler is written in Scala and runs on the JVM, whereas Verilator is written in C++.
\figRef{compile_time} contains a detailed breakdown of compilation steps and their contribution to the total time.
Most of the compile-time is spent in parallelizing RTL code.
We expect to improve compilation time (it is currently unoptimized).

\subsection{FPGA resource utilization}\label{sec:utilization_appendix}

\tabRef{utilization} reports FPGA resource utilization for a single core.
The dominant resource is the URAM, as two are required per core (one for the instruction memory and one for the local scratchpad).
This limits the total number of cores on the U200 to 398.
% However, as discussed in the evaluation, a 225-element grid is usually the breaking point in scaling (see~\figRef{manticore_scaling}).
As all cores may not use their scratchpad memory, one optimization is a heterogeneous implementation where some cores lack a scratchpad and rely on only a large register file so that other cores can have more local memory.
We leave heterogeneous processor grids to future work.

\begin{lstlisting}[language=verilog,
    % xleftmargin=2em,
    % xrightmargin=2em,
    numberstyle=\scriptsize\ttfamily,
    captionpos=b,
    numbers=left,
    stepnumber=1,
    numbersep=1.5pt,
    float=h,
    caption={Simple Verilog Example.}, label={lst:verilog_example}
    ]
module EvenOdd(input wire clock);
  reg [15:0] counter = 0;
  always @(posedge clock) begin
    counter <= counter + 1;
    if (counter[0] == 1'b0) begin
        $display("%d is an even number", counter);
    end else begin
        $display("%d is an odd number", counter);
    end
    if (counter == 20) begin
        $finish;
    end
  end
endmodule
\end{lstlisting}

\begin{lstlisting}[language=masm,
  caption={Manticore assembly code for \codeRef{verilog_example}.}, label={lst:masm_example},
  float=h
]
.p0: // privileged process
    // init $r0 = 0, $r1 = 1, $r3 = 0, $r5 = 0, $r6 = 0
    PRED $r1  // predicate mask for global store
    GST  $r4, cat{$r0, $r0, $r0} // global store $r2 in address 0
    // if ($r6) $display("%d is an even number", $r4);
    EXPECT  $r6, $r0, 0
    // if ($r6) $display("%d is an odd number", $r4);
    EXPECT  $r5, $r0, 1
    EXPECT  $r3, $r0, 2 // if ($r3) $finish
    // recv $r5, $r3, $r4 from p1
    // sleep
    // implicit jump  to p0

.p1:
    // init $r0 = 0, $r1 = 1, $r2 = 20, $r4 = 0
    // $r4 is the counter
    SLICE  $r3, $r4[0 +: 1] // $r3 = counter[0]
    SEQ	 $r5, $r4, $r2 // $r5 = (counter == 20)
    ADD	 $r7, $r4, $r1 // $r7 = counter + 1
    SEND p0.$r4, $r4 // p0.$r4 = counter;
    NOP  7;
    MUX  $r8, $r3, $r1, $r0 // $r8 = !counter[0]
    SEND p0.$r5, $r3 // p0.$r5 = counter[0] (delayed)
    SEND p0.$r3, $r5 // p0.$r3 = (counter == 0)
    MOV $r4, $r7     // counter = counter + 1
    NOP  7;
    SEND p0.$r6, $r8
    // implicit jump to p1
\end{lstlisting}
% \begin{figure}
%   \centering
%   \subfloat[Example Verilog]{
%     \lstinputlisting[language=verilog,
%       xleftmargin=-1em,
%       % xrightmargin=2em,
%       numberstyle=\scriptsize\ttfamily,
%       captionpos=b,
%       % numbers=left,
%       % stepnumber=1,
%       % numbersep=3pt,
%       % float=htpb
%       %caption={A Verilog counter.}, label={lst:verilog_example}
%       ]{listing/even_odd.v}
%   }\\
%   \subfloat[Equivalent in Manticore assembly]{
%     \lstinputlisting[language=masm
%       %caption={Manticore assembly code for \codeRef{verilog_example}}, label={lst:masm_example}]
%     ]{listing/p0.masm}
%   }
%   \caption{Translating a Verilog counter to Manticore assembly.}
%   \label{fig:verilog_example}
% \end{figure}

\begin{table}
  \centering
  \resizebox{1\columnwidth}{!}{%
    \begin{tabular}{c c c c c c c c}
      \toprule
      \textbf{}   & \textbf{LUT} & \textbf{LUTRAM} & \textbf{FF} & \textbf{BRAM} & \textbf{URAM} & \textbf{DSP} & \textbf{SRL} \\
      \midrule
      \#          & 545          & 128             & 1358        & 4             & 2             & 1            & 102          \\
      \textbf{\%} & 0.05         & 0.02            & 0.05        & 0.19          & 0.21          & 0.01         & 0.02         \\
      \bottomrule
    \end{tabular}
  }
  \caption{
    Resource utilization of a single core on the U200.
    Percentages are the faction of the total resources available on an U200.
  }
  \label{tab:utilization}
\end{table}

\begin{table}
  \centering
  \setlength{\tabcolsep}{1.5pt}
  \resizebox{\linewidth}{!}{%
    \begin{tabular}{l c  c c c c c }
      \toprule
      \multirow{2}{*}{\textbf{Bench}} & \multirow{2}{*}{$|E|$} & \multirow{2}{*}{$|V|$} & \multirow{2}{*}{\textbf{LoC}} & \multicolumn{3}{c}{\textbf{Compile Time}}                                              \\
      \cmidrule{6-7}
                                      &                        &                        &                               & \textbf{Manticore}                        & \textbf{Verilator} & \textbf{Verilator MT} \\
      \midrule
      \textbf{vta}                    & 56142                  & 7037                   & 190818                        & 15m29s                                    & 2m33               & 26s                   \\
      \textbf{mc}                     & 52330                  & 9182                   & 30353                         & 12m57s                                    & 1m13s              & 16s                   \\
      \textbf{noc}                    & 114364                 & 6927                   & 39363                         & 15m14s                                    & 3m23s              & 36s                   \\
      \textbf{mm}                     & 89102                  & 6659                   & 64963                         & 8m38s                                     & 7m5s               & 2m55s                 \\
      \textbf{rv32r}                  & 60430                  & 4497                   & 31761                         & 5m57s                                     & 1m56s              & 29s                   \\
      \textbf{cgra}                   & 57532                  & 4615                   & 104498                        & 7m48s                                     & 2m15s              & 37s                   \\
      \textbf{bc}                     & 8135                   & 4630                   & 276                           & 2m23s                                     & 40s                & 27s                   \\
      \textbf{blur}                   & 9649                   & 751                    & 3869                          & 42s                                       & 22s                & 15s                   \\
      \textbf{jpeg}                   & 1005                   & 131                    & 6542                          & 16s                                       & 7s                 & 3s                    \\
      \bottomrule
    \end{tabular}
  }
  \caption{
    Manticore, single-thread compile Verilator and multithreaded compile Verilator compilation times.
    $|E|$ and $|V|$ respectively denote the number of edges and nodes in the graph obtained by splitting each benchmark into a maximal set of independent processes (see \secRef{parallelization}).
    \textbf{LoC} denotes the Verilog lines of code for each benchmark.
  }
  \label{tab:compile_time}
\end{table}

% \subsection{Parallel Simulation Modeling}

\begin{figure*}
    \centering
    \subfloat[Core i7-9700K]{\includegraphics[width=0.25\textwidth]{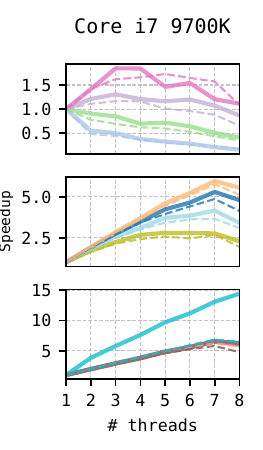}}
    \subfloat[EPYC 7V73X]{\includegraphics[width=0.75\textwidth]{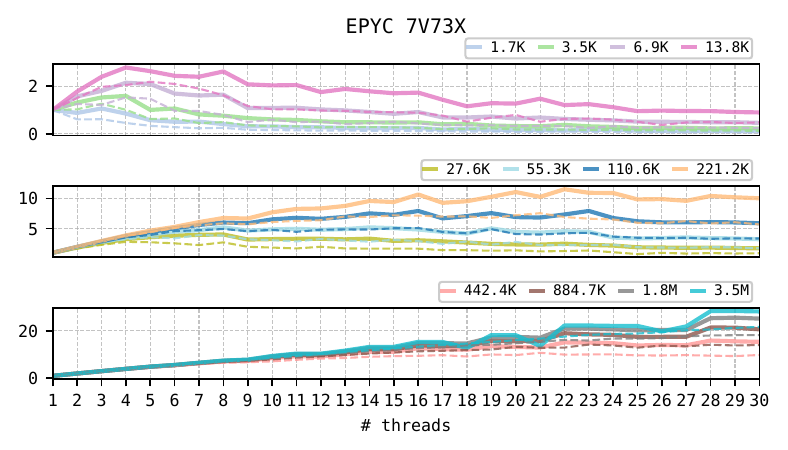}}
    \vspace{-10pt}
    \subfloat{
        \resizebox{1\textwidth}{!}{%
            \begin{tabular}{ l l cccccccccccc  }
                \toprule
                \multicolumn{2}{l}{\bf Instructions }     & \textbf{1.7K}  & \textbf{3.5K} & \textbf{6.9K} & \textbf{13.8K} & \textbf{27.6K} & \textbf{55.3K} & \textbf{110.6K} & \textbf{221.2K} & \textbf{442.4K} & \textbf{884.7K} & \textbf{1.8M} & \textbf{3.5M} \\
                \midrule
                \multirow{2}{*}{\rotatebox{90}{\bf epyc}} & {\bf model 1 } &
                [381, 6569]                               &
                [375, 3824]                               &
                [385, 2512]                               &
                [302, 1804]                               &
                [322, 1164]                               &
                [209, 820]                                &
                [105, 533]                                &
                [52, 396]                                 &
                [26, 281]                                 &
                [13, 186]                                 &
                [7, 119]                                  &
                [3, 71]                                                                                                                                                                                                                                               \\
                                                          & {\bf model 2 } &
                [368, 3558]                               &
                [372, 2665]                               &
                [382, 1796]                               &
                [375, 1166]                               &
                [209, 845]                                &
                [105, 545]                                &
                [52, 415]                                 &
                [26, 302]                                 &
                [13, 208]                                 &
                [7, 140]                                  &
                [3, 84]                                   &
                [2, 46]                                                                                                                                                                                                                                               \\
                \midrule
                \multirow{2}{*}{\rotatebox{90}{\bf i7}}   & {\bf model 1 } &
                [1509, 8821]                              &
                [1527, 4411]                              &
                [1438, 2568]                              &
                [1106, 1913]                              &
                [555, 1485]                               &
                [277, 1016]                               &
                [139, 677]                                &
                [69, 400]                                 &
                [35, 218]                                 &
                [17, 100]                                 &
                [9, 58]                                   &
                [4, 30]                                                                                                                                                                                                                                               \\

                                                          & {\bf model 2 } &
                [1134, 7336]                              &
                [1509, 3718]                              &
                [1610, 2429]                              &
                [916, 1692]                               &
                [464, 1299]                               &
                [232, 965]                                &
                [116,613]                                 &
                [58, 345]                                 &
                [29, 189]                                 &
                [14, 94]                                  &
                [7,    46]                                &
                [2,    26]                                                                                                                                                                                                                                            \\
                \bottomrule
            \end{tabular}
        }
    }
    \label{fig:fake_rate_appendix}
    \caption{Measured parallel simulation speedup on a desktop processor (left) and server processor (right). %
        Dashed lines model only synchronization cost. Solid lines also include i-cache pressure. %
        Each curve is labeled by the number of instructions executed per simulation step. %
        The table at the bottom, shows the [min., max.] simulation rates corresponding to each model.
    }
    % \end{subfigure}
\end{figure*}

% \end{figure}

% \begin{figure*}
%   \lstinputlisting[language=c++, caption={A simple model of parallel RTL simulation without modeling frontend bottlenecks}, label={lst:first_order_model}]{listing/fake_rate.cpp}
% \end{figure*}

% \begin{sidewaysfigure*}
%   \includegraphics[width=\linewidth,angle=0]{figures/verilator_dyn_instr.pdf}
%   \caption{
%     Verilator performance on an Intel Xeon E5-2680 v3.
%     Breakdown of CPU empty pipeline slots on the left axis and executed instruction (normalized to single-thread) on the right axis.
%   }
%   \label{fig:verilator_dyn_instr}
% \end{sidewaysfigure*}

% \begin{sidewaysfigure*}
%   \includegraphics[width=\linewidth]{figures/microarch.pdf}
%   \caption{
%     Microarchitecture of a core in Manticore's processor grid. Details are omitted for legibility.
%   }
%   \label{fig:manticore_micro}
% \end{sidewaysfigure*}

\end{document}